\documentclass{aa}
\usepackage{graphicx} 
\usepackage{txfonts}
\usepackage{amsmath}

\usepackage{natbib,twoopt}
\usepackage{hyperref}

\usepackage{enumitem}
\usepackage{float}

\usepackage{placeins}
\usepackage{longtable}
\usepackage{lscape}

\usepackage{threeparttable}

\usepackage{placeins}
\usepackage{array}
\usepackage{subcaption}
\usepackage{dcolumn}

\bibpunct{(}{)}{;}{a}{}{,}             

\hypersetup{
  colorlinks=true, 
  urlcolor=blue, 
  linkcolor=blue, 
  citecolor=blue 
}

\newcounter{Rco}
\newcommand{\Ionst}[1]{\setcounter{Rco}{#1}\Roman{Rco}}

\newcommand{\Ionw}[3]{\mbox{#1\,{\scriptsize\Ionst{#2}}~{$\lambda\,#3$}\,\AA}}

\newcommand{\Ionww}[3]{\mbox{#1\,{\scriptsize\Ionst{#2}}~$\lambda\lambda\,#3$\,\AA}}

\newcommand{\teff}{\textit{T}\textsubscript{eff}\xspace}
\newcommand{\logg}{\mbox{log \textit{g}}\xspace}

\newcommand{\teffpm}[2]{\mbox{\teff = #1 $\pm$ #2 kK}}
\newcommand{\loggpm}[2]{\mbox{\logg = #1 $\pm$ #2}}

\newcommand{\Msol}{\ensuremath{M_\odot}}

\newcommand{\Mdot}{\ensuremath{\dot{M}}\xspace}

\newcommand{\Mloss}{\mbox{\Msol}yr\textsuperscript{-1}\xspace}

\newcommand{\logNfracE}[2]{\mbox{log (#1/H)\xspace= \xspace#2}}

\newcommand{\comp}[1]{\textsuperscript{(#1)}}
\newcommand{\comptwo}[2]{\textsuperscript{(#1)}\textsuperscript{,}\xspace\textsuperscript{(#2)}}

\newcommand{\comptwoA}[2]{\textsuperscript{(#1)}\textsuperscript{,}\xspace\textsuperscript{(#2)}\textsuperscript{,}}

\newcommand{\compthr}[3]{\textsuperscript{(#1)}\textsuperscript{,}\xspace\textsuperscript{(#2)}\textsuperscript{,}\xspace\textsuperscript{(#3)}}

\newcommand{\compthrA}[3]{\textsuperscript{(#1)}\textsuperscript{,}\xspace\textsuperscript{(#2)}\textsuperscript{,}\xspace\textsuperscript{(#3)}\textsuperscript{,}}

\newcommand{\up}{\textsuperscript{$\triangledown$}}

\begin{document}

   \title{Spectral evolution of hot hybrid white dwarfs \\ I. Spectral analysis}

   \author{Semih Filiz\inst{\ref{inst1}}
          \and
          Klaus Werner\inst{\ref{inst1}}
          \and
          Thomas Rauch\inst{\ref{inst1}}
          \and
          Nicole Reindl\inst{\ref{inst2}}}
   \institute{Institut f\"ur Astronomie und Astrophysik, Kepler Center for Astro and Particle Physics, Eberhard Karls Universit\"at, Sand 1, 72076 T\"ubingen, Germany \\
            \email{filiz@astro.uni-tuebingen.de}\label{inst1}
          \and
          Landessternwarte Heidelberg, Zentrum f\"ur Astronomie, Ruprecht-Karls-Universit\"at, K\"onigstuhl 12, 69117 Heidelberg, Germany \label{inst2}
            }
            
    \date{Received 14 August, 2024; accepted 3 October, 2024}

 
  \abstract
   {Hydrogen-rich white dwarfs (WDs) comprise the majority of the WD population, but are only rarely found at the very hot end of the WD cooling sequence. A small subgroup that exhibits both hydrogen and helium lines in their spectra, the so-called hybrid (or DAO) WDs, represents the majority of hydrogen-rich WDs at effective temperatures \teff $\approx$\,100\,kK.} 
   {We aim to understand the spectral evolution of hot hybrid WDs. Although small in number, they represent an evolutionary phase for most ($\approx$\,75 \%) WDs.}
   {We conducted a nonlocal thermodynamic equilibrium (NLTE) analysis with fully metal line blanketed model atmospheres for the ultraviolet (UV) and optical spectra of a sample of 19 DA and 13 DAO WDs with \teff $>$\,60\,kK. The UV spectra allow us to precisely measure the temperature through model fits to metal lines in different ionization stages. This enables us to place the WDs accurately on the cooling sequence.}
   {In contrast to earlier studies that typically relied on temperature measurements made from hydrogen lines alone, all DAOs in our sample are clearly hotter than the DAs. DAOs transform into DAs when they cool to \teff $\approx$\,75--85\,kK, depending on their mass. Along the cooling sequence, we witness a gradual decrease in the abundance of helium and the CNO elements in the DAOs due to gravitational settling. Simultaneously, iron and nickel abundances increase up to the transition region because radiative forces act more efficiently on them. This is followed by a steady decline. We discuss the implications of our results on atomic diffusion theory and on the role of weak radiation-driven winds in hot hydrogen-rich WDs.}
   {}


    \maketitle
\nolinenumbers

\section{Introduction}
\label{sec:Intro}
White dwarfs (WDs) represent the final stage in the evolution of low- and intermediate-mass stars (M $< 8$ \mbox{\Msol}). They are the predominant end products of stellar evolution. From the \emph{Gaia} mission, approximately 359\,000 high-confidence WD candidates were identified in the Milky Way \citep{2021MNRAS.508.3877G}, and more than 37\,000 WDs are spectroscopically confirmed \citep[and references therein]{2021MNRAS.507.4646K}. With the upcoming space-based gravitational-wave detector Laser Interferometer Space Antenna (LISA), optically faint WD binaries will be studied. This in turn will help us to uncover Galaxy regions that are otherwise hard to observe \citep{2019MNRAS.490.5888L}.
 
In addition to their substantial numbers, WDs present an incontrovertible prospect of investigating physics under extreme conditions. In this sense, hot WDs offer a unique opportunity besides connecting the final stages of stellar evolution to earlier phases. They are reliable sources for deriving atomic data of elements beyond the iron group \citep{2012A&A...546A..55R,2014A&A...564A..41R,2014A&A...566A..10R,2015A&A...577A..88R,2015A&A...577A...6R,2016A&A...587A..39R,2016A&A...590A.128R,2017A&A...599A.142R,2017A&A...606A.105R,2020A&A...637A...4R} and for testing theories of the influence of gravity on the fundamental constants \citep{2021MNRAS.500.1466H}.

In the aftermath of the asymptotic giant branch (AGB) phase, the post-AGB star enters the WD cooling sequence with either an H-rich (spectral type DA) or He-rich (spectral type DO) envelope of the main constituent. About three-quarters of all WDs are born H-rich \citep{2020ApJ...901...93B}, but along the cooling track, this fraction does not remain constant because atmospheric processes alter the main spectroscopic marker \citep{2022ApJ...927..128B}. As in this case, the drastic change in the atmospheric composition is referred to as spectral evolution. In the course of cooling off, WD atmospheres are prone to transformation by multiple processes, such as mass loss, radiative levitation, gravitational settling, atomic diffusion, convection, and accretion. Our current understanding of the spectral evolution of WDs has been established through the collective efforts of nearly four decades, and addressing every milestone in this endeavor is beyond the scope of this introduction. Therefore, we only mention the required aspects and developments for hot H-rich and hybrid WDs (spectral type DAO) that show a weak \Ionw{He}{2}{4686} line. For an extensive review of the spectral evolution of WDs, we refer to \citet{2024Ap&SS.369...43B}.

The presence of helium in the hot hydrogen-rich WD atmospheres has been the subject of several discussions. Extreme-ultraviolet (EUV) and X-ray observations of hot DA WDs revealed a flux deficiency in the shorter wavelengths \citep{1988ApJ...331..876V}, which is opposed to what is expected for hot WDs with pure H envelopes. It was initially proposed that traces of He in the atmosphere provide the opacity for the observed EUV and X-ray flux deficiency. However, \citet{1988ApJ...331..876V} demonstrated that the required amount of He could not be radiatively levitated. They showed that EUV observations can instead be reproduced with models that consider stratified atmospheres in which a thin layer of H (\textit{M}\textsubscript{H}/\ensuremath{M_\ast} = 10$\textsuperscript{-15}$--10$\textsuperscript{-13}$) floats above the He envelope. However, this contradicts standard stellar evolution theory, which predicts \textit{M}\textsubscript{H}/\ensuremath{M_\ast} = 10$\textsuperscript{-4}$ for the H envelope \citep{1997ASSL..214...57B}. Subsequent studies showed that the EUV and X-ray fluxes of several other hot DA WDs cannot be reproduced with stratified atmospheres \citep{1989ApJ...336L..25V,1992ApJ...390..590V,1992ApJ...401..288V}. Metal opacities were instead identified as the likely source of the observed flux deficiency in these objects.

The discovery of hybrid WDs \citep{1979A&A....71..163K} helped to clarify the situation. \citet{1994ApJ...432..305B} demonstrated that models with homogeneous atmospheres yielded better fits than stratified atmospheres in the optical spectra of DAO WDs (except for one object\footnote{This object is an example for DAO WDs with stratified atmospheres, which are found at \teff $\leq$ 55 kK \citep{2016ApJ...833..127M,2020ApJ...901...93B} and advance through the DO-to-DA evolutionary channel \citep{2016ApJ...833..127M}.} at that time). Since the radiation pressure is proven insufficient to prevent the gravitational settling of He \citep{1988ApJ...331..876V}, they suggested that weak mass-loss might induce the homogenized hybrid atmospheres. \citet{1994ApJ...432..305B} also reported that their analysis was heavily hindered by the Balmer-line problem (BLP), which commonly emerges in the optical spectra of hot DA and DAO WDs and prevents simultaneous fits to all Balmer lines with the same parameters. The authors concluded that the BLP is linked to the presence of metals in the atmosphere, which reinforced the idea that metal-driven winds shape the atmospheres of hot hybrid WDs. The BLP and the trace-metal connection was later supported by the quantitative and qualitative spectral analyses of \citet{1996ApJ...457L..39W} and \citet{2010ApJ...720..581G}, respectively. On the other hand, the origin of the metals in hot WD atmospheres has been heavily debated because of the mismatch between observations and diffusion theory that assumes an equilibrium between gravitational settling and radiative levitation \citep{1995ApJS...99..189C,1995ApJ...454..429C}. Mass loss was mainly regarded as the missing ingredient in theoretical diffusion calculations. Only \citet{2014MNRAS.440.1607B} and \citet{2019MNRAS.487.3470P} speculated that the accretion of planetary material might explain the observed metal abundances.

The occurrence of stellar winds in hot WDs cannot be shown directly because the expected mass-loss rates are too low \citep{2000A&A...359.1042U} for a spectroscopic detection in P Cygni profiles. Nonetheless, indications of mass loss can still be deduced from the spectra of hot WDs. The first indication for the occurrence of winds was the detection of ultra-highly excited (UHE) metals in the optical spectra of hot WDs \citep{1995A&A...293L..75W}. Because of the asymmetrical shapes of the line profiles, it was speculated that UHE features stem from shock fronts in the stellar wind. \citet{2019MNRAS.482L..93R} showed that recurrent spectroscopic variations of UHE lines and the light curve can be explained with rotational modulation of a magnetosphere that traps the ejected material. In this case, magnetically confined wind shocks can heat the plasma to the required extreme temperatures ($10\textsuperscript{6}$ K) for the occurrence of the UHE phenomenon. Another indirect evidence for radiation-driven winds was shown by \citet{1999A&A...350..101N}. They reported a decreasing He abundance with decreasing luminosity in their hybrid WD sample.

Theoretical calculations of the influence of radiation-driven winds on the atmospheric composition by \citet{1998A&A...338...75U,2000A&A...359.1042U} revealed that mass-loss rates of $\approx$\,\mbox{$10\textsuperscript{-11}$ \Mloss} can retard the gravitational settling of He in the hybrid atmospheres. The inevitable cessation of the mass loss expedites the effects of gravity, and a spectral transition from DAO to DA is expected. This paints a clear picture in the Kiel diagram, which has a well-defined theoretical wind limit. However, observations of hot DA and DAO WDs showed a lower effective temperature (\teff) range for the transition \citep{2010ApJ...720..581G}. Additionally, the measured temperature and surface gravity (\logg) values of both H-rich and hybrid WDs revealed that DA and DAO WDs partially overlap along the cooling sequence \citep{1994ApJ...432..305B,1999A&A...350..101N,2003MNRAS.341..870B,2004MNRAS.355.1031G,2010ApJ...720..581G,2011ApJ...743..138G}. When we accept that the mass loss in hybrid WDs is strong enough to prevent He from sinking out of the atmosphere, then we have to ask why the winds of DAs with similar \teff and \logg are insufficient to sustain He in the atmosphere. This indicates that other factor(s) in addition to \teff and \logg are involved in inducing the spectral transformation along the cooling track. The scatter of hot DA and DAOs in the \teff-\logg diagram was indeed speculated to be a consequence of the initial metallicity or of different masses \citep{2000A&A...359.1042U,2020ApJ...901...93B}. Therefore, our immediate aim is to measure the element abundances in UV spectra of hot DA and DAO WDs in order to determine the atmospheric and stellar parameters. We also wish to shed light on whether a well-defined limit exists in the Kiel diagram that separates DA and DAO WDs, and if the metal abundances change along the cooling track.

In this paper, we analyze the ultraviolet (UV) and optical spectra of DA and DAO WDs\footnote{Conventionally, the WD spectral classification is based on evaluated optical spectra. In this paper, we exclude secondary letters denoting weaker spectral features detected in other parts of the electromagnetic spectrum for simplicity (e.g., Z in the case of a metal detection for DAZ and DAOZ WDs).}. In Sect. \ref{sec:Obs}, the sample selection and observations are described. We explain the spectral analysis procedure for measuring atmospheric and stellar parameters in Sects. \ref{sec:SpecAnalysis} and \ref{sec:Masses}. Finally, we discuss the implications of our results in Sect. \ref{sec:discussion} and give a brief summary in Sect. \ref{sec:summary}.

\section{Sample selection and observations}
\label{sec:Obs}
We selected a sample of hot WDs with \teff $\geq$ \mbox{60 kK}, above which homogeneous hybrid WDs are found. Our sample comprised 13 DAO, and 19 DA WDs (Table \ref{tab:Observations}) for which the main selection criterion was the availability of archival Far Ultraviolet Spectroscopic Explorer (FUSE) or Space Telescope Imaging Spectrograph (STIS) spectra. All objects have archival FUSE spectra, except for WD\,0939+262. Archival STIS spectra are present only for 4 DAOs and 8 DAs.

\begin{figure*}[h!]
\centering
  \includegraphics[width=18cm,trim=0cm 8.5cm 0cm 0.5cm, clip]{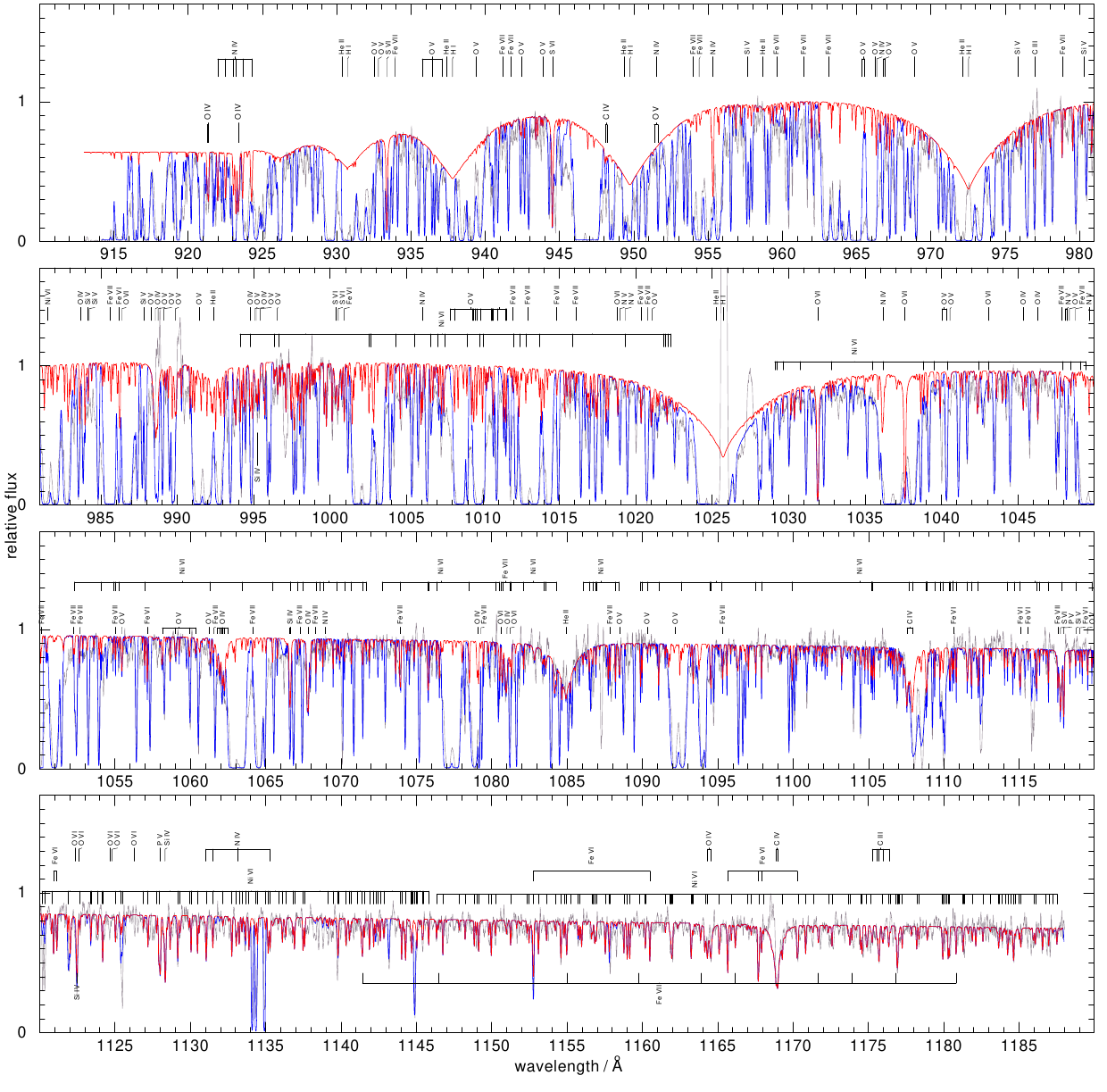}
    \caption{FUSE spectrum (gray) of the DAO WD\,2342+806 (red shows the model with \teff = \mbox{83 kK}, \logg = 7.2 (cm/s$^2$), and blue shows the same model including the ISM lines). All model spectra displayed in this paper are convolved with Gaussians according to the instrument resolution. Additionally, all UV spectra are smoothed with a low-bandpass filter.}      \label{fig:WD2342+806_FUSE}
\end{figure*}

\begin{figure*}[h!]
\centering
  \includegraphics[width=18cm,trim=0cm 8.5cm 0cm 0.5cm, clip]{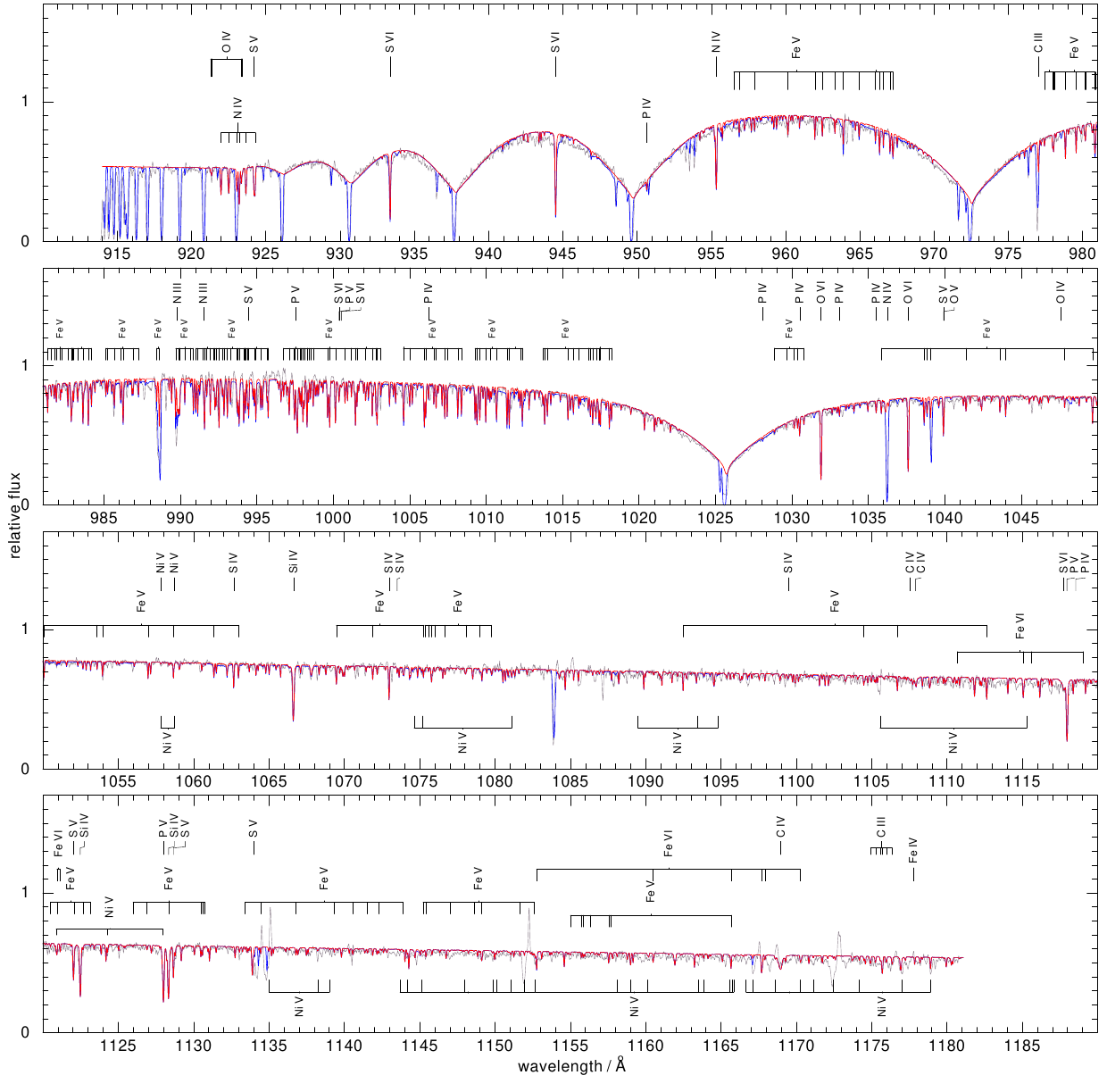}
    \caption{FUSE spectrum of the DA WD\,2211$-$495 (red shows the model with \teff = \mbox{68 kK}, \logg = 7.4, and blue shows the model including the ISM lines).}
    \label{fig:WD2211-495_FUSE}
\end{figure*}

\subsection{FUSE}
\label{subsec:FUSE_obs}
We retrieved FUSE spectra for 31 objects observed with the low-resolution spectrograph (LWRS; resolving power \mbox{\emph{R} $\approx$\,20\,000}) aperture from Mikulski Archive for Space Telescopes (MAST)\footnote{\url{https://archive.stsci.edu}}. The entire FUSE wavelength range (905--1188 {\AA}) consists of eight overlapping segments. Therefore, an inspection of flux levels from each segment as well as each exposure was necessary. When any exposure-to-exposure flux variation exceeded \mbox{$\approx$\,10\%}, we discarded the corresponding exposure. The majority of exposures made with the LiF1b channel suffer from a flux decrease because the detector area is shadowed by grid wires, known as the "worm" (FUSE Data Handbook\footnote{\url{https://archive.stsci.edu/fuse/instrumenthandbook/contents.html}}). When the overall flux level of a particular LiF1b spectrum did not drastically deviate from the spectra of other channels, the wavelength range in which the worms prevailed was eliminated. Otherwise, the entire LiF1b spectrum was discarded, which was our action in most cases. Then, each available exposure from a particular segment was coadded into a single spectrum. Moreover, the coadded spectrum of each overlapping segment was then coaligned to already identified ISM lines and combined into a single composite spectrum. Finalized composite FUSE spectra of DAO WD\,2342+806 and DA WD\,2211-495 with their best-fit models are shown as examples in Figs. \ref{fig:WD2342+806_FUSE} and \ref{fig:WD2211-495_FUSE}, respectively.

\subsection{HST}
\label{subsec:HST_obs}
The archival spectra of 12 sample objects were observed with the STIS on board the \emph{Hubble} Space Telescope (HST) using the FUV-MAMA detector. For 9 of the objects (4 DAO and 5 DA), observations were carried out with the E140M grating ($\approx$\,1144--1729 {\AA}, \emph{R} $\approx$\,45\,800), and the remainder were observed with the E140H grating (\mbox{\emph{R} $\approx$\,114\,000}), which consists of prime and secondary tilts. In the MAST archive, observations made with E140H only contain prime tilt with a central wavelength of 1416 {\AA}, meaning that we only had access to the spectral range of 1317--1517 {\AA} for these 3 DAs. When more than one observation was available, the observations were coadded to increase the signal-to-noise ratio (S/N). In Fig.~\ref{fig:WD2342+806_HST} and \ref{fig:WD2211-495_HST}, STIS spectra of WD\,2342+806 (E140M) and WD\,2211$-$495 (E140H), respectively, are shown as examples with their best-fit models.

\subsection{Optical}
\label{subsec:Opt_obs}
Archival optical spectra for all sample objects were acquired from the Montreal White Dwarf Database \citep[MWDD,][]{2017ASPC..509....3D}, except for Longmore 1, WD\,0851+090, WD\,1111+552, and WD\,2350$-$706. The typical resolution for the MWDD spectra is \mbox{$\approx$\,3--6 {\AA}}. Observations of the MWDD spectra are described in detail by \citet{2010ApJ...720..581G,2011ApJ...743..138G}, \citet{2011AJ....141...96L}. We found no optical spectrum of WD\,1111+552 and WD\,2350$-$706, but archival Ultraviolet and Visual Echelle Spectrograph (UVES; \mbox{\emph{R} $\approx$\,40\,000}) spectra are available for \mbox{Longmore 1} and WD\,0851+090. UVES spectra of WD\,0621$-$376 and WD\,2211$-$495 were also used to make comparisons. For the same reason, we acquired Sloan Digital Sky Survey \citep[SDSS,][]{2009ApJS..182..543A} spectra of three other DAs (WD\,1056+516, WD\,1342+443, WD\,1827+778), which were also available in MWDD. We should note that for nine other objects (two DAOs and seven DAs) UVES and (four DAOs and five DAs), LAMOST spectra are available. However, conducting an extensive optical analysis is beyond the scope of this study. Therefore, we simply analyzed optical spectra to constrain \logg and investigated whether common ground between optical and UV results can be found.

\subsection{Interstellar absorption and reddening}
\label{subsec:ISM_redenning}
Absorption lines from the interstellar medium (ISM) may heavily contaminate FUSE spectra, and they need to be disentangled from photospheric lines to precisely determine the atmospheric parameters. Therefore, we identified ISM lines employing the line-fitting procedure OWENS \citep{2002ApJS..140...67L,2002P&SS...50.1169H,2003ApJ...599..297H}, which individually models radial and turbulent velocities, column densities, temperatures, and chemical compositions of different clouds. In general, the detected ISM lines included \ion{D}{i}, \ion{H}{i}, H\textsubscript{2}, \ion{C}{I-III}, \ion{N}{I-III}, \ion{O}{I}, \ion{P}{II}, \ion{S}{II-III}, \ion{Ar}{I}, and \ion{Fe}{II}. 

We determined the interstellar reddening for each star by comparing the model fluxes to Tycho2, SDSS, and 2MASS magnitudes. The model spectra were normalized to fluxes observed with the filter in the longest wavelength when no 2MASS K flux was available. Then, the reddening law by \citet{1999PASP..111...63F} with \textit{R}\textsubscript{V} = 3.1 was employed to determine the \textit{E(B-V)} values of the sample objects.

\section{Spectral analysis}
\label{sec:SpecAnalysis}

We conducted a spectral analysis with the Tübingen Model-Atmosphere package \citep[TMAP\footnote{\url{http://astro.uni-tuebingen.de/~TMAP}},][]{1999JCoAM.109...65W, 2003ASPC..288...31W,2012ascl.soft12015W}, which computes nonlocal thermodynamic equilibrium (NLTE) model atmospheres in radiative and hydrostatic equilibrium assuming plane-parallel geometry. The opacities of H, He, C, N, O, Si, P, S, Fe, and Ni were included in our models. Classical model atoms (hydrogen to sulfur) were constructed with atomic data from the Tübingen Model Atom Database \citep[TMAD,][]{2003ASPC..288..103R}. For Fe and Ni, we used the Tübingen Iron-group opacity interface \citep[TIRO,][]{2003ASPC..288..103R,2013PhDT.......668M}, which uses a statistical approach to manage the vast number of atomic levels and line transitions by combining energy levels and sampled lines of a particular ion to super levels and super lines, respectively. Additionally, we employed the Stark broadening tables of \citet{2009ApJ...696.1755T} and \citet{1989A&AS...78...51S} to calculate the synthetic line profiles of \ion{H}{i} and \ion{He}{ii}.

\begin{table}
    \caption{Number of NLTE levels and lines of extended classical model ions.}
    \renewcommand*{\arraystretch}{1.25}
    \centering
    \begin{tabular}{c c c c c c c}
    \hline\hline
     &\ion{}{ii}&\ion{}{iii}& \ion{}{iv}& \ion{}{v} & \ion{}{vi} \\
    \hline
    C   &       1, 0    &       56, 242 &       54, 295 &                   &          \\
    N   &       1, 0    &       34, 129 &       90, 547 &       54, 297 &             \\
    O   &       1, 0    &       72, 322 &       83, 637 &  103, 729     & 54, 291 \\
    Si  &                   &   17, 27  &       30, 102 &       25, 59  & 27, 74  \\
    P   &                   &   3, 1    &       21, 25  &       18, 49  &             \\
    S   &                   &   1, 0    &       37, 150 &       39, 107 & 25, 48  \\ 
    \hline   
    \end{tabular}
    \\
    \smallskip
     {\raggedright \small \textbf{Notes.} Each cell in the table shows the number of NLTE levels and lines. The highest ionization stage of each element is only included as ground state and is not listed here. \par}
    
    \label{tab:levels}
\end{table}

Metal line blanketed models were initially built with relatively small model atoms for elements up to sulfur to avoid numerical instabilities and to optimize computation time. Subsequently, line formation iterations were executed with extended model atoms by keeping the atmospheric structure fixed. An overview of the extended classical model atoms is listed in Table \ref{tab:levels}. Additionally, ionization stages \ion{}{IV} - \ion{}{ix} were included for Fe and Ni (Table \ref{tab:FeNi_levels}). We customized model atoms for each object by adjusting the ionization stages of a particular element according to individual parameter ranges. 

\begin{table}
\caption{Number of super levels and super lines of the iron and nickel model atoms.}
\renewcommand*{\arraystretch}{1.15}
    \centering
    \begin{tabular}{l c r r}
    \hline\hline
    &Super Levels & Super Lines & Sample Lines  \\
    \hline
    \ion{Fe}{IV}   &    7       &       25      &       3102371 \\
    \ion{Fe}{V}    &    7       &       25      &       3266247 \\
    \ion{Fe}{VI}   &    7       &       25      &       991935  \\
    \ion{Fe}{VII}  &    7       &       24      &       200455  \\
    \ion{Fe}{VIII} &    7       &       27      &       19587   \\
    \ion{Fe}{IX}   &    1       &       0       &       0       \\
    \ion{Ni}{IV}   &    7       &       25      &       2512561 \\
    \ion{Ni}{V}    &    7       &       27      &       2766664 \\
    \ion{Ni}{VI}   &    7       &       27      &       7407763 \\
    \ion{Ni}{VII}  &    7       &       25      &       4195381 \\
    \ion{Ni}{VIII} &    7       &       27      &       1473122 \\
    \ion{Ni}{IX}   &    1       &       0       &       0       \\
        \hline
        \end{tabular}
    \label{tab:FeNi_levels}
\end{table}

The wide range in temperature and surface gravity and the numerous parameters of our sample did not facilitate either creating an extensive model atmosphere grid or taking a statistical approach in the spectral analysis. Instead, our method involved an iterative process of line-profile fitting by consecutively computing a series of models while improving the parameters and decreasing the uncertainty. Our analysis procedure is summarized below.  

\begin{enumerate}
    \item Calculate a small grid of pure H and H+He models close to the literature \teff and \logg values of DA and DAOs, respectively.
    \item Roughly constrain \teff, \logg, and He abundance from the UV H and He lines.
    \item Compute a series of fully metal-line blanketed models around the updated values.
    \item Refine \teff and determine the metal abundances by fitting the UV metal lines.
    \item Improve \logg and He abundance from the optical spectra.
    \item Recalculate models with the fine-tuned parameters.
\end{enumerate}

\subsection{DAO analyses}
\label{subsec:SpecDAO}

\subsubsection{Effective temperature and surface gravity}
\label{subsubsec:SpecDAOtefflogg}

\begin{figure*}[h!]
\centering
  \includegraphics[width=17.5cm,trim=1cm 6cm 1.1cm 1cm, clip]{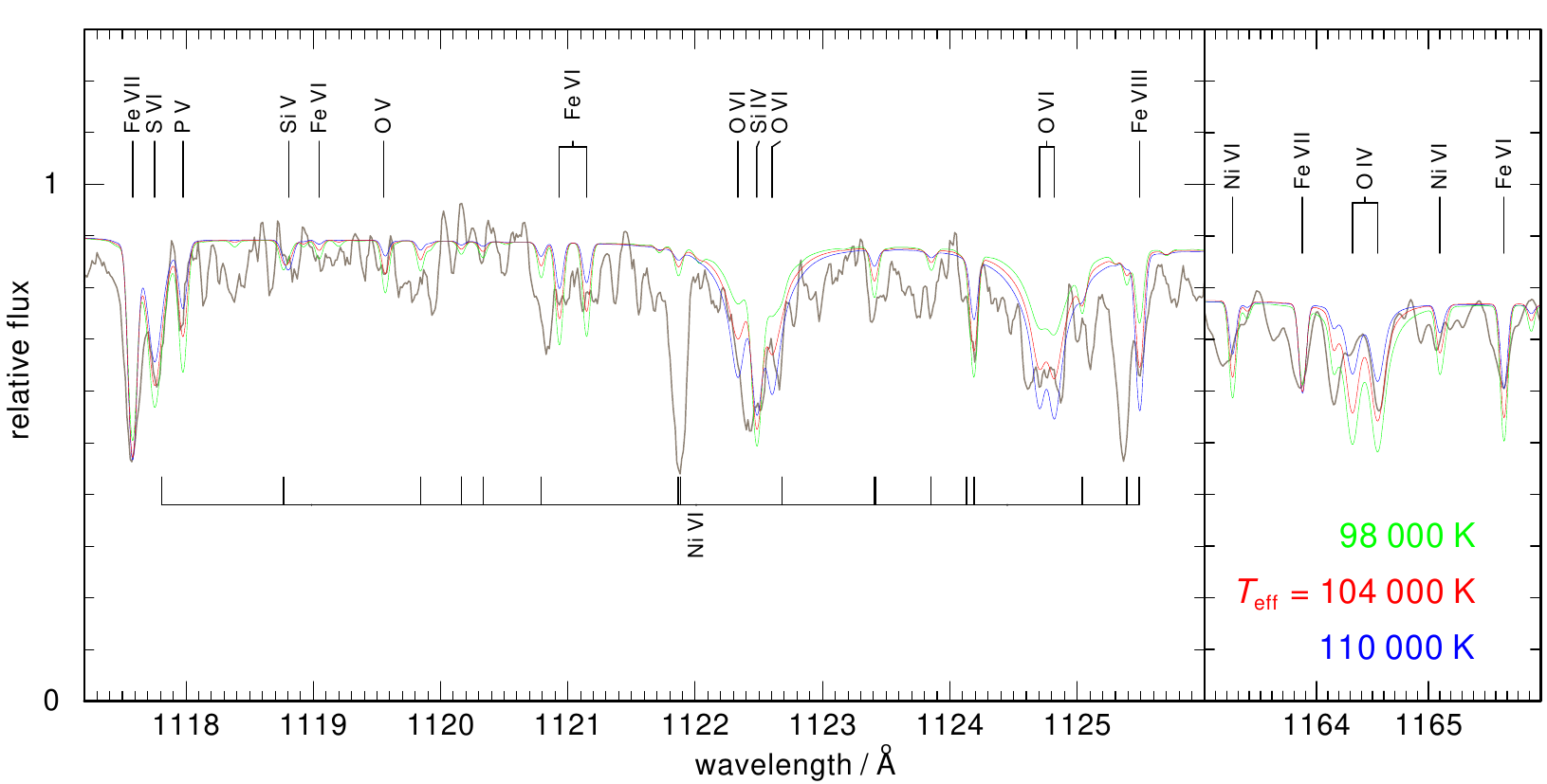}
    \caption{Details of the FUSE spectrum of the DAO WD\,0500$-$156 compared to three models with different temperatures \teff = 104 $\pm$ 6 kK, \logg = 7.2.}
    \label{fig:WD0500-156_Fe_O}
\end{figure*}

H+He models for each object were computed assuming a solar He content, with \teff and log \textit{g} close to the values determined by \citet{2010ApJ...720..581G} through their analysis of optical spectra via Balmer-line fitting. When the latter authors did not analyze a particular object, we selected values close to those derived by \citet{2012PhDT.......152Z}. Then, we constrained \teff and log \textit{g} from the Lyman lines in the FUSE spectra. In the next step, we included metal opacities in our model atoms. A set of strategic lines from several ions was identified since the absence or presence of a particular line would indicate a certain \teff. While keeping log \textit{g} fixed, we exploited multiple ionization equilibria such as \ion{C}{iii}/\ion{C}{iv}, \ion{N}{iv}/\ion{N}{v}, \ion{O}{iv}/\ion{O}{v}/\ion{O}{vi}, \ion{Fe}{v}/\ion{Fe}{vi}/\ion{Fe}{vii}/\ion{Fe}{viii}, and \ion{Ni}{v}/\ion{Ni}{vi} to constrain \teff (Fig.~\ref{fig:WD0500-156_Fe_O}). A detailed overview of each element is presented in Sect. \ref{subsec:SpecDAO_Abund}.

Contamination due to interstellar \ion{H}{i} and other interstellar absorption hinders the correct assessment of the local continuum in the far-UV range. This in turn affects the precise measurement of \logg via fitting the Lyman lines. Thus, investigation of optical spectra was necessary to verify our \logg values. In general, the parameters determined with UV analysis lay within the error range of the Balmer-line analysis.

\subsubsection{He abundance}
\label{subsubsec:SpecDAOHe}
A preliminary He abundance was determined from isolated lines of \ion{He}{ii} \mbox{\textit{n} = 2 $\rightarrow$ \textit{n}'} line series in the FUSE range. While we fine-tuned the temperature and gravity, the He abundance was further adjusted accordingly. Finally, we compared and, if necessary, modified our previous measurement by using \Ionw{He}{2}{1640} (\mbox{\textit{n} - \textit{n}' = 2 - 3}) and \Ionw{He}{2}{4686}, which are the optimal lines to adjust the He abundance in their respective wavelength regions.

\subsubsection{Metal abundances}
\label{subsec:SpecDAO_Abund}

\begin{description}[wide,itemindent=\labelsep]

\item[Carbon.] 
The \Ionww{C}{4}{1107, 1169} doublet was mainly used to determine the C abundance. In several cases, \Ionw{C}{4}{1169} is contaminated by an airglow line. When this was the case, we only relied on \Ionw{C}{4}{1107}. Although weak \ion{C}{iii} features can be observed below \mbox{100 kK}, only WD\,2342+806 displays the \Ionw{C}{3}{1175} multiplet, which is a blend with \ion{N}{vi} lines. The STIS wavelength range contains \ion{C}{iv} multiplets at 1198 and 1239 {\AA} as well as the \ion{C}{iv} resonance doublet. Without exception, the line fits of the former two agree with the other C lines from the FUSE spectra, whereas the latter is weaker in our models. However, a quick inspection revealed that this must be due to ISM contribution. We achieved a good fit to the line cores of the photospheric components, which are clearly separated from the blueshifted ISM component. 
\smallskip

\item[Nitrogen.] 
The relative strengths of identified \ion{N}{iv} and \ion{N}{v} lines in the FUSE range can be exploited to confine \teff within a rather wide range. The \Ionw{N}{4}{923} multiplet and the \Ionw{N}{4}{955} singlet disappear around \mbox{120 kK}, whereas \Ionw{N}{5}{1048.2} and \Ionw{N}{5}{1049.7} quickly become weaker below \mbox{90 kK}. The latter two are blended with ISM lines and cannot be identified in our sample. Therefore, the \Ionw{N}{4}{923} multiplet and the \Ionw{N}{4}{955} singlet were mainly used to determine the N abundance. When STIS spectra were available, the \Ionww{N}{5}{1238.2, 1242.8} resonance doublet and the \Ionw{N}{4}{1718.5} were used as well. Compared to our photospheric models, we encountered a stronger \ion{N}{v} resonance doublet in the observations of WD\,1957+225 and WD\,2226$-$210, which is expected to be dominated by interstellar absorption.
\smallskip

\item[Oxygen.] 
Multiple \ion{O}{iv}, \ion{O}{v}, and \ion{O}{vi} lines are detectable in the FUSE spectra. A rather tight constraint can be made with the O ionization balance since O lines strongly react to \teff changes. A weak \ion{O}{vi} doublet at 1124 {\AA} becomes visible around \mbox{95 kK} and becomes stronger with increasing \teff, in contrast to the \ion{O}{iv} lines (Fig.~\ref{fig:WD0500-156_Fe_O}). At \mbox{110 kK}, the \ion{O}{V} lines in the FUSE wavelength range still display strong features (Fig.~\ref{fig:Oion_WD1111}). However, above this temperature and below \mbox{90 kK}, they diminish quickly. We encountered problems with the \Ionww{O}{6}{1031.9, 1037.6} resonance doublet of the objects with \teff below 115 kK. Without exception, the line cores are too deep in our final models. This issue was also reported by \citet{2007A&A...470..317R} and \citet{2018A&A...616A..73W}. In the STIS wavelength range, \Ionw{O}{5}{1371.7} and \Ionww{O}{4}{1338.6, 1342.9, 1343.5} were identified. STIS spectra of WD\,1957+225 and WD\,2226$-$210 display \Ionw{O}{6}{1291}. However, in both cases, the observations show very broad features, and we were unable to model this line well.
\smallskip

\begin{figure}[]
\resizebox{\hsize}{!}{\includegraphics[trim=0.5cm 5.7cm 5cm 10cm, clip]{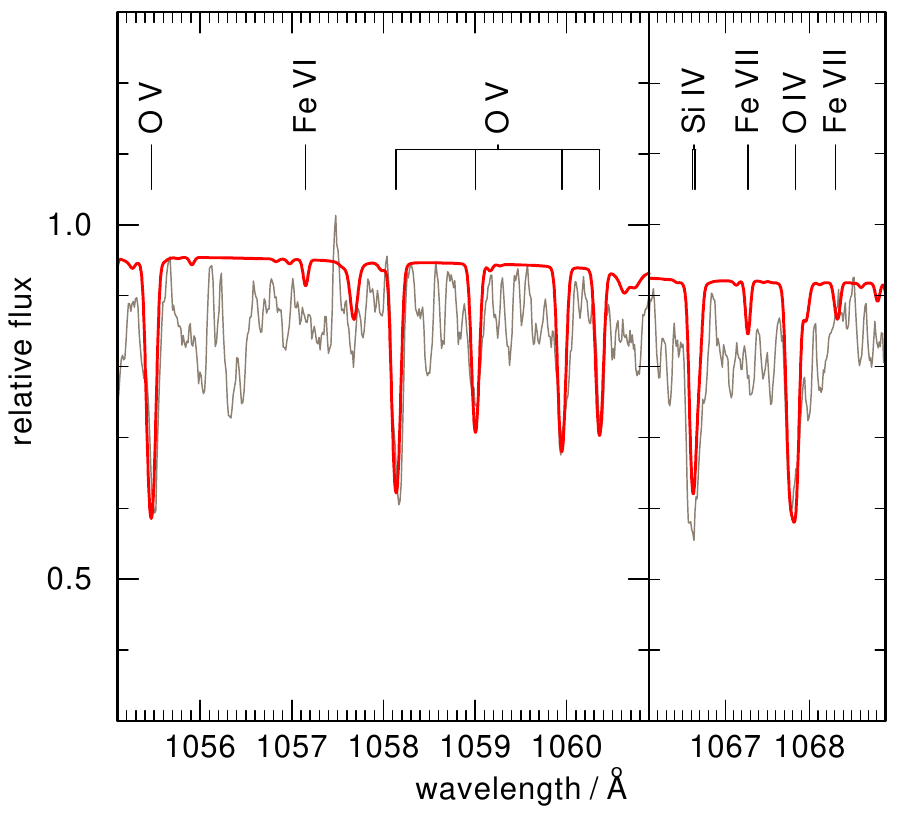}}
  \caption{Sections of the FUSE spectrum (gray) of the DAO WD\,1111+552 showing the ionization balance of \ion{O}{iv} and \ion{O}{v} (red shows the model with \mbox{\teff = 111 kK}, \logg = 7.1).}
    \label{fig:Oion_WD1111}
\end{figure}

\item[Silicon.]
The most prominent silicon lines are \Ionww{Si}{4}{1066.61, 1066.65} and \Ionww{Si}{4}{1122,1128} in the FUSE range. However, \Ionw{Si}{4}{1066} is blended with interstellar absorption. Therefore, the \Ionww{Si}{4}{1122,1128}  doublet was used to determine the Si abundance. Only three objects with \teff above \mbox{110 kK} display a weak \Ionw{Si}{5}{1118.8} feature. Additionally, STIS spectra show the \Ionww{Si}{4}{1394, 1402} resonance doublet, which is dominated by interstellar absorption, like the other two resonance lines (\ion{C}{IV}, \ion{N}{V}) identified in the STIS spectra. As opposed to the FUSE spectra, we detected \ion{Si}{V} lines (\Ionww{Si}{5}{1245.7, 1251.4, 1276}) above \mbox{97 kK}, which become stronger with increasing \teff.

\item[Phosphorus.]
The P abundances were determined by fitting the \Ionww{P}{5}{1118, 1128} resonance doublet. The other detected \ion{P}{V} lines are quite weak and blend with other lines. Our DAOs are too hot to show \ion{P}{IV} lines. No phosphorus lines were detected in the STIS spectra.

\item[Sulfur.]
We identified the \Ionww{S}{6}{933, 945} resonance doublet, \Ionw{S}{6}{1000}, and \Ionw{S}{6}{1117.7} in the FUSE spectra. Only one object resides in the temperature range in which \ion{S}{V} lines can be displayed. Only \Ionw{S}{6}{1419.4, 1419.7} and \Ionw{S}{6}{1423.8} were detected in the STIS spectra of WD\,0439+466 and WD\,2342+806.

\begin{figure}[]
\resizebox{\hsize}{!}{\includegraphics[trim=0.7cm 6cm 6.3cm 10cm, clip]{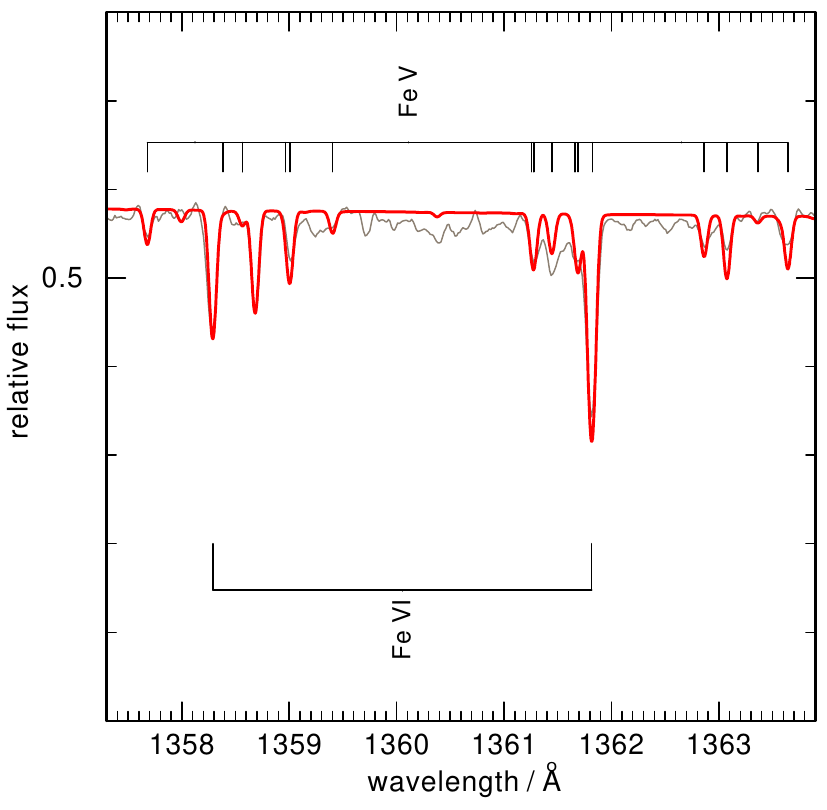}}
  \caption{Section of the STIS spectrum (gray) of the DAO WD\,0439+466  (red shows the model with \mbox{\teff = 97 kK}, \logg = 7.0).}
    \label{fig:FeV_WD0439}
\end{figure}

\item[Iron.]
Prominent \ion{Fe}{vi}\ion{ -}{viii} lines were detected in the FUSE spectra, and they serve as excellent \teff indicators. In Fig.~\ref{fig:WD0500-156_Fe_O}, the \teff assessment of WD\,0500-156 is depicted as an example. As opposed to \ion{Fe}{vi}, the relative strength of the \ion{Fe}{viii} lines intensifies above \mbox{98 kK}, whereas the increase is minor for the \ion{Fe}{vii} lines. At \mbox{110 kK}, the \ion{Fe}{viii} lines become too strong, while the \ion{Fe}{vi} lines weaken substantially. In addition to the mentioned ions, \ion{Fe}{v} lines were also identified in the STIS spectra of objects below 100 kK (Fig.~\ref{fig:FeV_WD0439}).

\begin{figure}[]
\resizebox{\hsize}{!}{\includegraphics[trim=0.6cm 6cm 6.1cm 10cm, clip]{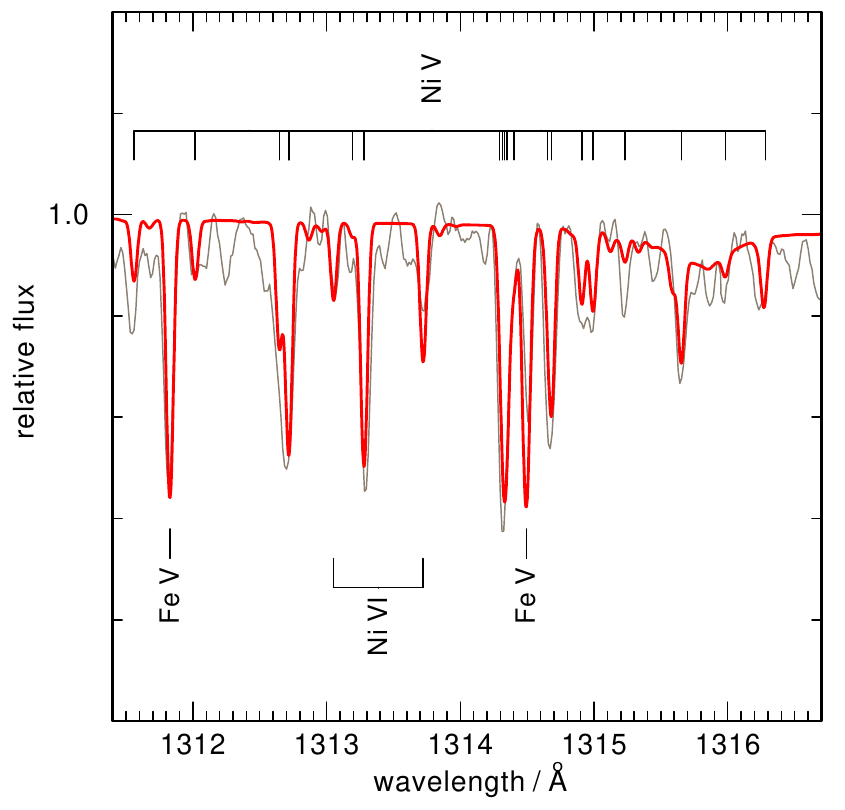}}
  \caption{Detail of the STIS spectrum (gray) of the DAO WD\,2342+806  (red shows the model with \mbox{\teff = 83 kK}, \logg = 7.2).}
    \label{fig:NiV_WD2342}
\end{figure}

\smallskip

\item[Nickel.]
FUSE spectra contain many \ion{Ni}{vi} lines. The line strength of \ion{Ni}{vi} lines diminishes quickly above \mbox{100 kK} (Fig.~\ref{fig:WD0500-156_Fe_O}). In the STIS spectra of WD\,0439+466 and WD\,2342+806, a number \ion{Ni}{v} and \ion{Ni}{vi} lines were identified (Fig.~\ref{fig:NiV_WD2342}).

\end{description}

\subsection{DA analyses}
\label{subsec:Spec_DA}

\begin{figure*}[ht!]
\centering
  \includegraphics[width=17.5cm,trim=0.6cm 6cm 0cm 4cm, clip]{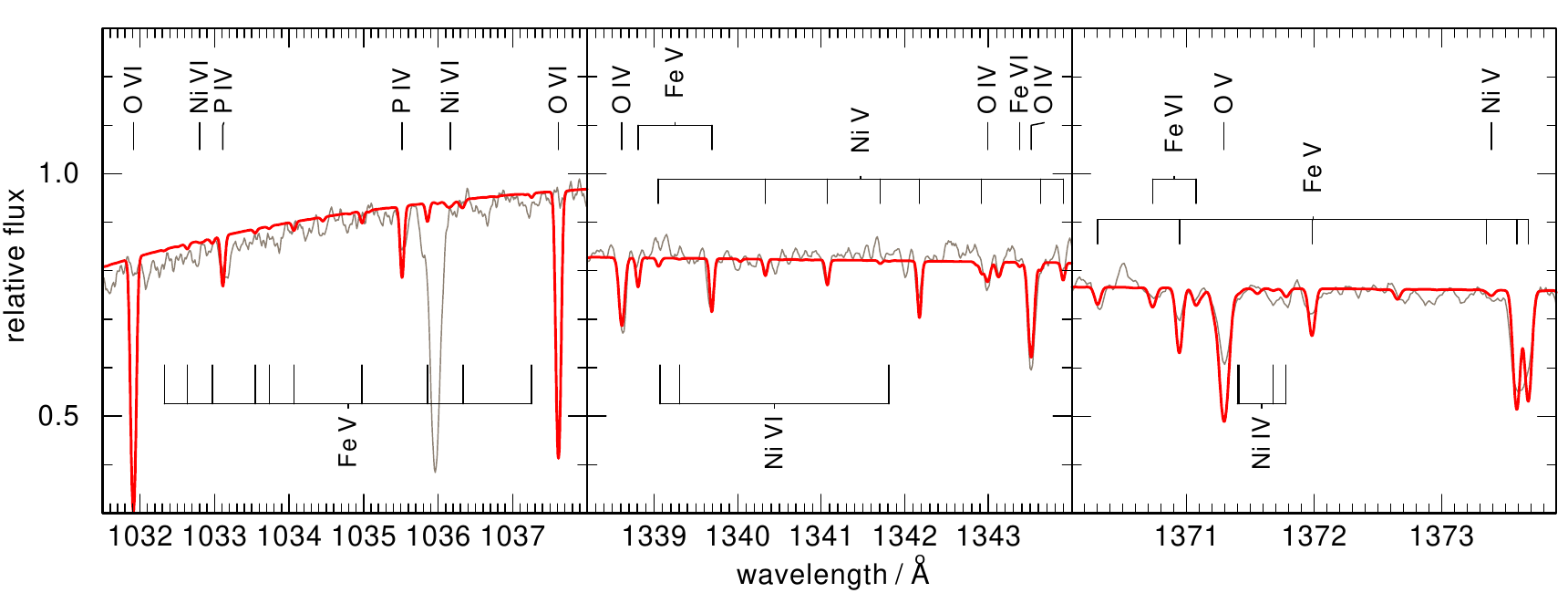}
    \caption{Sections of FUSE (left panel) and STIS (middle and right panels) spectra (gray) of the DA WD\,0232+035 (red shows the model with \mbox{\teff = 63 kK}, \logg = 7.5), illustrating stratification effects on oxygen ions.}
    \label{fig:O_inbalance_WD0232}
\end{figure*}

\subsubsection{Effective temperature and surface gravity}
\label{subsubsec:Spec_DA_tefflogg}
We followed the same procedure as for the DAOs, but started by computing pure H models with literature values of \teff and \logg \citep{2010ApJ...720..581G,2011ApJ...743..138G,2014MNRAS.440.1607B}. In addition to adding the metal opacities, we also included a small amount of He in our model atmospheres. In their far-UV analysis of G191-B2B, \citet{2013A&A...560A.106R} opted for log He < $-4.7$ (mass fraction) as an upper limit. Since G191-B2B resides at the lower limit of the temperature criterion in our sample selection, this value was chosen as our starting He abundance and was adjusted to upper limits from \ion{He}{ii} \mbox{\textit{n} - \textit{n}' = 2 - 11} line series for each object. Additionally, we were able to fine-tune the upper limit with \Ionw{He}{2}{1640} for the objects with STIS spectra. Since the DAs in our sample have considerably lower temperatures than the DAOs, the ionization equilibria that we relied on also slightly altered (\ion{C}{iii}/\ion{C}{iv}, \ion{N}{iii}/\ion{N}{iv}/\ion{N}{v}, \ion{O}{iv}/\ion{O}{v}/\ion{O}{vi}, \ion{P}{iv}/\ion{P}{v}, \ion{S}{iv}/\ion{S}{v}/\ion{S}{vi}, \ion{Fe}{v}/\ion{Fe}{vi}/\ion{Fe}{vii}, and \ion{Ni}{v}/\ion{Ni}{vi}). Because of the low metal abundance and poor data quality, it was not possible to use the ionization balance of multiple elements in some cases. For these objects, higher error limits were imposed. In contrast to the DAO WDs, \logg values determined utilizing Lyman lines were not compatible with the Balmer lines. Therefore, we readjusted \logg according to the Balmer lines. However, in some cases, the disparity between the Lyman and the lower-order Balmer lines (e.g., H\,$\alpha$ and H\,$\beta$) was too large, but the surface gravity determined from Lyman lines completely agreed with the higher-order Balmer lines. Since fitting higher-order Balmer lines corresponds to a more accurate \logg measurement \citep{1996ApJ...457L..39W} and \teff could be tightly constrained from UV, we opted in these cases for \logg measured in the UV and adopted a larger error margin.

\subsubsection{Metal abundances}
\label{subsubsec:Spec_DA_abund_O_stra}

\begin{description}[wide,itemindent=\labelsep]
\item[Carbon.] 
The \Ionw{C}{3}{1175} multiplet was mainly used to determine C abundances. Multiple objects show a weak \Ionw{C}{4}{1169} doublet. Similar to DAOs, this feature is contaminated in several FUSE spectra. When HST spectra were available in addition to \ion{C}{iv} multiplets at 1198 and 1239 {\AA}, we used the \Ionww{C}{4}{1548,1550} resonance doublet to confirm the C abundance. 
\smallskip

\item[Nitrogen.]
Most of the sample objects display the \ion{N}{iv} multiplet at 923 {\AA} and the singlet at 955 {\AA}. \Ionw{N}{3}{991} was also identified in the spectra of objects with \teff $\leq$ \mbox{75 kK}. However, this line is blended with a \ion{Fe}{V} line, and it would be unrealistic to determine the N abundance with it, although it is still useful for assessing the temperature with the \ion{N}{iii}/\ion{N}{iv} ionization balance. Therefore, the former two lines were mainly used to determine the N abundance. When the STIS spectrum was present, we exploited the \Ionww{N}{5}{1238.2, 1242.8} resonance doublet as well.

\smallskip

\item[Oxygen.]
In the FUSE spectra of multiple objects, a weak \Ionww{O}{6}{1031.9, 1037.6} resonance doublet was identified. A handful of objects display the short-wavelength component of the \ion{O}{iv} 921--923 {\AA} multiple, but the other component is a blend with the \Ionw{N}{4}{923} multiplet. No additional oxygen lines were identified in the FUSE spectra. We detected \Ionw{O}{5}{1371.7} and \Ionww{O}{4}{1338.6, 1342.9, 1343.5} in the STIS spectra. However, it was not possible to achieve a simultaneous fit to the lines of all three ionization stages (see Fig.~\ref{fig:O_inbalance_WD0232}). We further discuss this in Sect. \ref{sec:discussion}.
\smallskip

\item[Silicon.]
The \Ionww{Si}{4}{1066.61, 1066.65}, and \Ionww{Si}{4}{1122,1128}  doublets were mainly used to determine the Si abundance, which agreed with fits to the \ion{Si}{iv} resonance lines. The \ion{Si}{iii} lines were identified.
\smallskip

\item[Phosphorus.]
The \Ionww{P}{5}{1118, 1128} resonance doublet was identified. Objects below $\approx$\,70 kK also display \Ionww{P}{4}{950,1028}. No other P lines were identified in the STIS spectra.
\smallskip

\item[Sulfur.]
Multiple cooler objects show the \Ionww{S}{4}{1072.97, 1073.52} doublet. This line quickly weakens above \mbox{70 kK}. In addition to \Ionw{S}{5}{1028}, \Ionw{S}{5}{1222}, and \Ionw{S}{5}{1501.7}, the \ion{S}{VI} resonance doublet was also identified. However, in multiple cases, we encountered a similar problem with the \ion{O}{vi} resonance doublet, in which the line cores were slightly stronger in our models. 
\smallskip

\item[Iron and nickel.]
Numerous \ion{Fe}{v} and \ion{Fe}{vi} as well as \ion{Ni}{v} and \ion{Ni}{vi} lines can be detected in FUSE and STIS spectra. The line strengths of both \ion{Fe}{vi} and \ion{Ni}{vi} start to diminish around \mbox{80 kK}, whereas \ion{Ni}{v} and \ion{Fe}{v} become stronger.
\smallskip

\end{description}

\section{Masses}
\label{sec:Masses}
\begin{figure}[h!]
\resizebox{\hsize}{!}{\includegraphics[trim=0cm 0cm 0cm 0cm, clip]{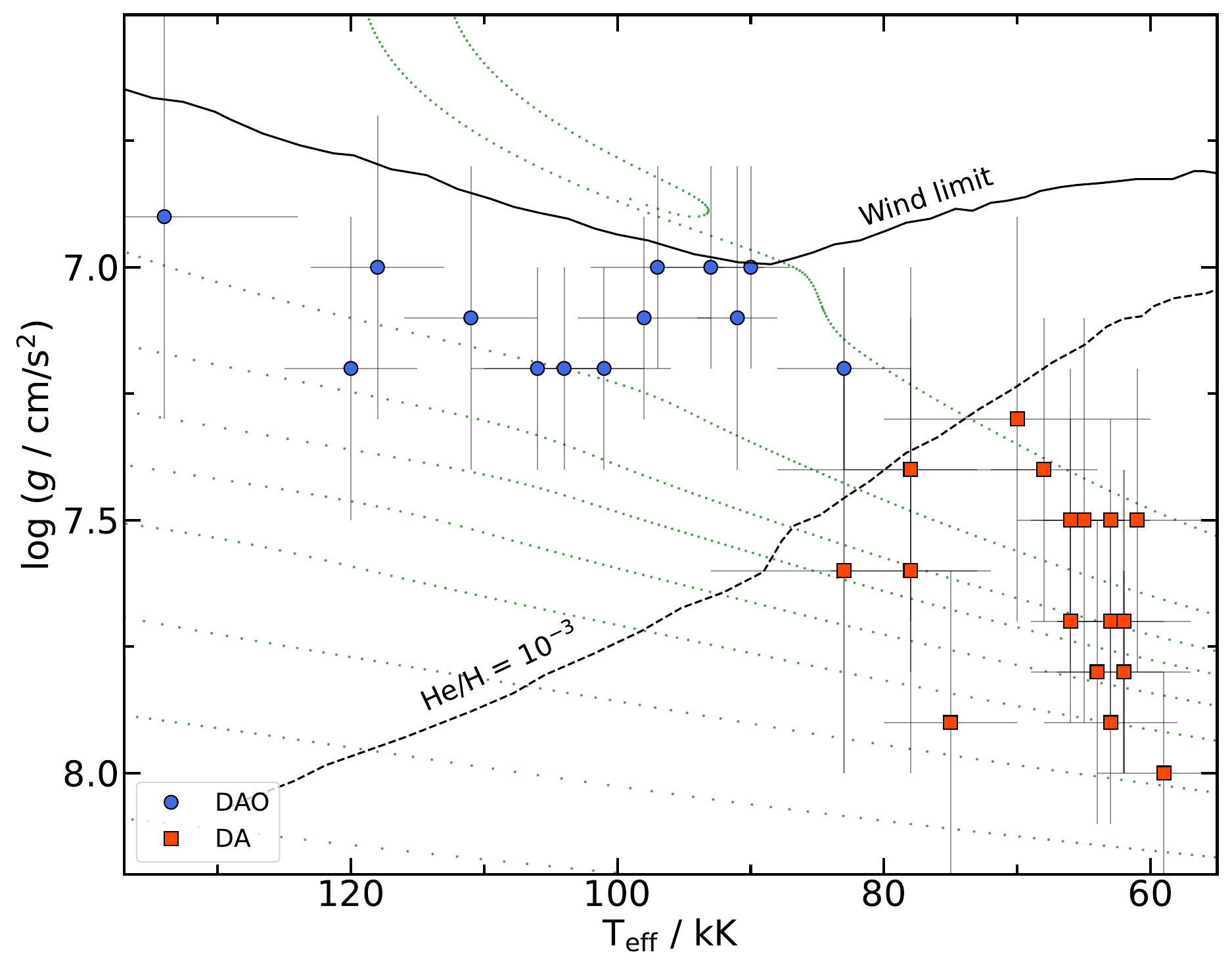}}
  \caption{Sample objects illustrated in the Kiel diagram. The dotted lines are evolutionary tracks of H-rich WDs (Z = 0.01) from \citet{2010ApJ...717..183R}, representing in order 0.525, 0.570, 0.593, 0.609, 0.632, 0.659, 0.705, 0.767, and 0.837 \Msol. The solid and dashed black lines correspond to the theoretical wind limit and to the He abundance (\mbox{\textit{N}(He)/\textit{N}(H) = 10$\textsuperscript{-3}$}) calculated with the predicted mass-loss rates by \citet[][see their Fig. 6]{2000A&A...359.1042U}, respectively. The latter also coincides with the approximate optical detection limit of He.}
    \label{fig:Kiel_diagram}
\end{figure}

We relied on the atmospheric parameters from our spectroscopic analysis to disclose the masses of the sample objects. To interpolate masses from the Kiel (\teff--\textit{g}) diagram, we followed the same steps as \citet{2023A&A...677A..29R} and used the griddata\footnote{\url{https://docs.scipy.org/doc/scipy/reference/generated/scipy.interpolate.griddata.html}} function in Python, which can rescale data points to the unit grid before making the interpolation. To estimate the Kiel masses, we considered the evolutionary tracks by \citet{2010ApJ...717..183R} and \citet{2013MNRAS.435.2048H} (Fig.~\ref{fig:Kiel_diagram}), which were devised for CO-core WDs (metallicity $Z = 0.01$) and He-core WDs, respectively. However, none of the object spectroscopic parameters match the He-core tracks. The uncertainties were assessed using a Monte Carlo method. We found that DAOs in our sample ($\langle$\textit{M}\textsubscript{DAO}$\rangle$ = 0.55 \Msol, $\sigma$ = 0.02 \Msol) are on average less massive than DAs ($\langle$\textit{M}\textsubscript{DA}$\rangle$ = 0.59 \Msol, $\sigma$ = 0.05 \Msol), which was also found by \citet{2010ApJ...720..581G}, \citet{2020ApJ...901...93B}, and \citet{2023A&A...677A..29R}.

\begin{figure*}[h!]
\centering
\includegraphics[width=18cm]{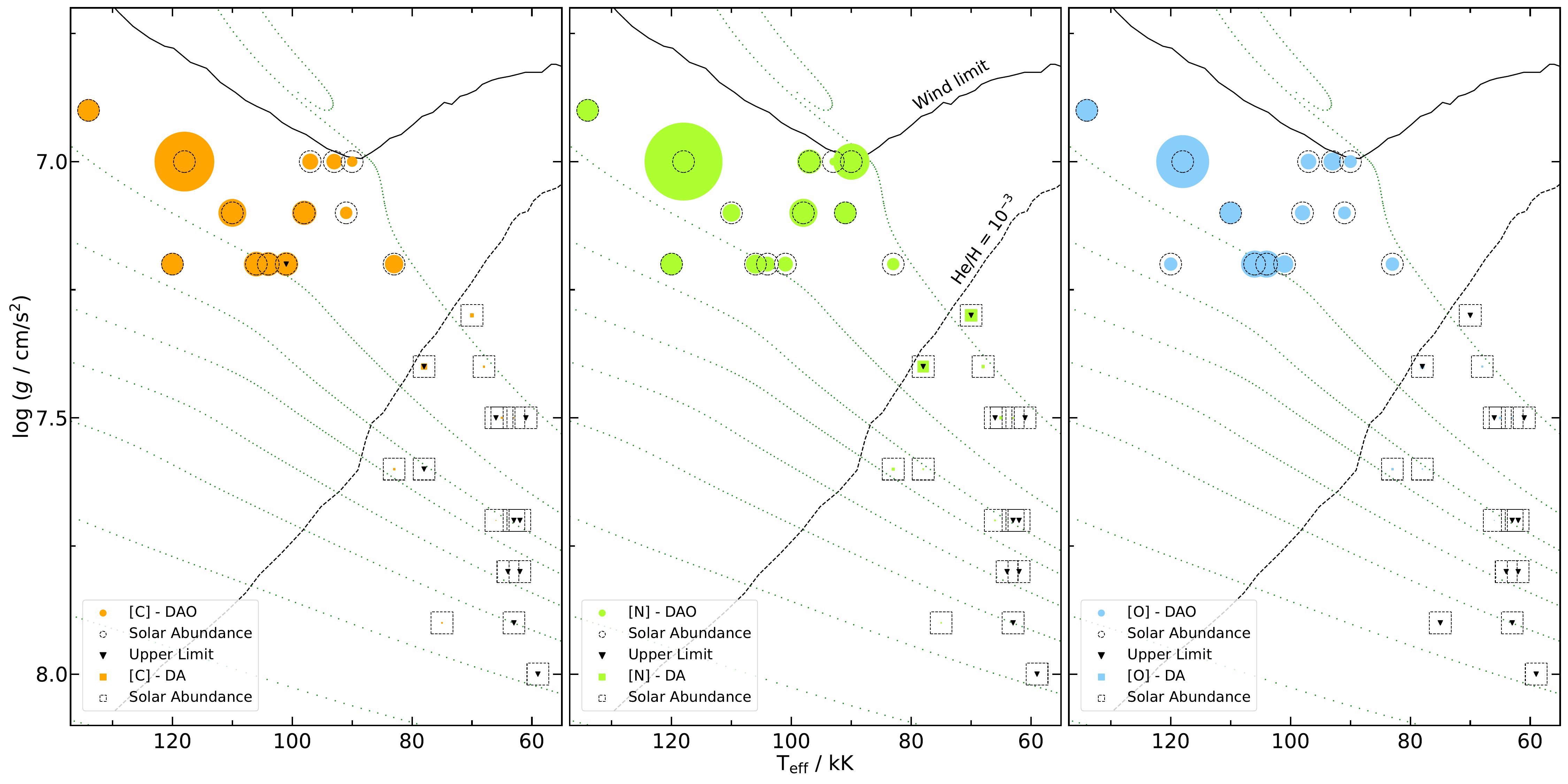}
\bigskip
\includegraphics[width=18cm]{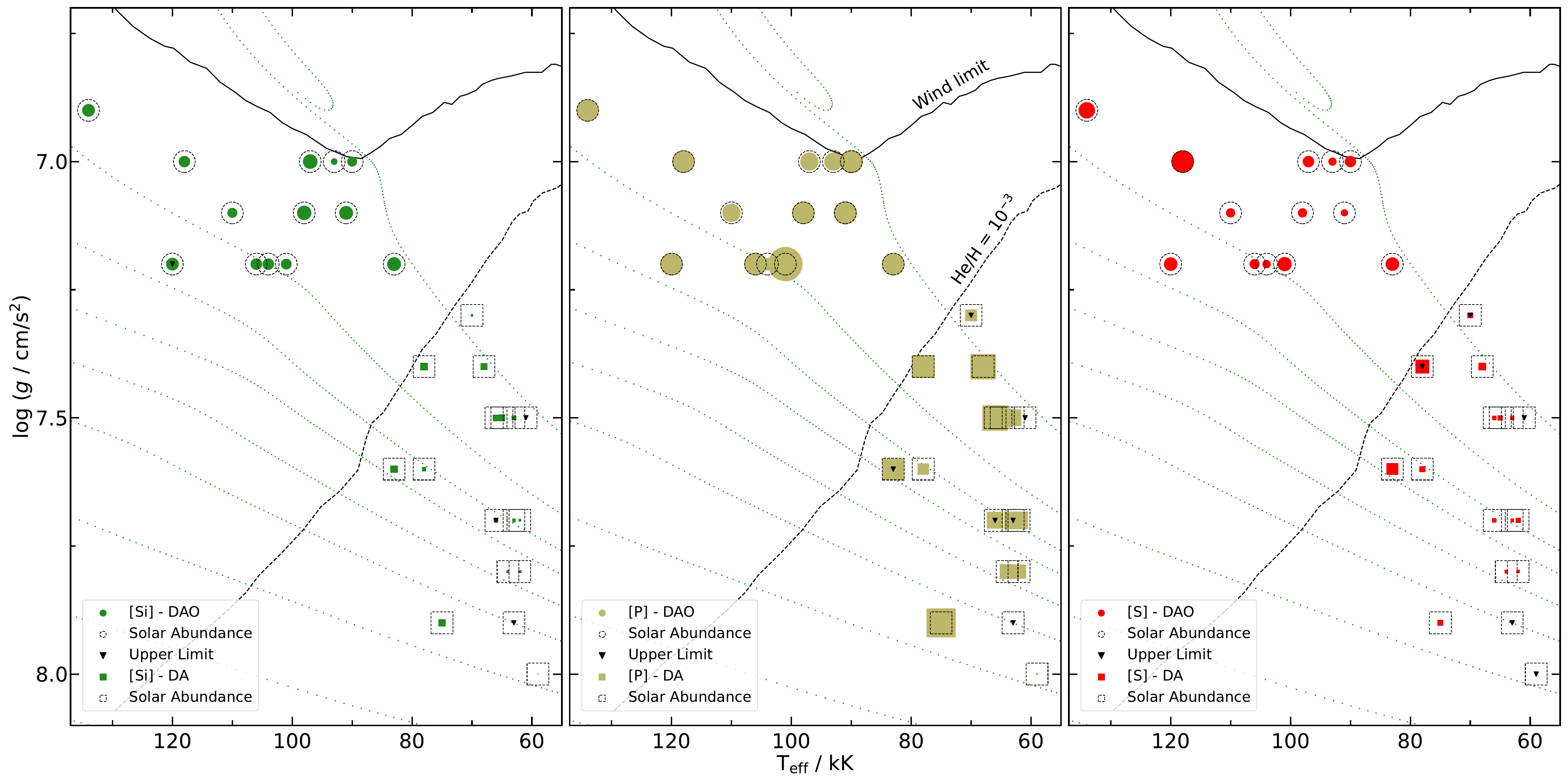}
\caption{Element abundances of our sample objects in the Kiel diagram (panels from upper left to lower right: C, N, O, Si, P, and S). The abundances are illustrated with filled circles and squares for DAOs and DAs, respectively. The symbol size is proportional to the mass fraction of the respective element. The dashed symbols represent solar values \citep{2009ARA&A..47..481A}. The upper limits are marked with an upside triangle, which does not scale with the abundances}.
\label{fig:CNO_Kiel}
\end{figure*}

\begin{figure*}[h!]
\centering
  \includegraphics[width=18cm,trim=0cm 0cm 0cm 0cm, clip]{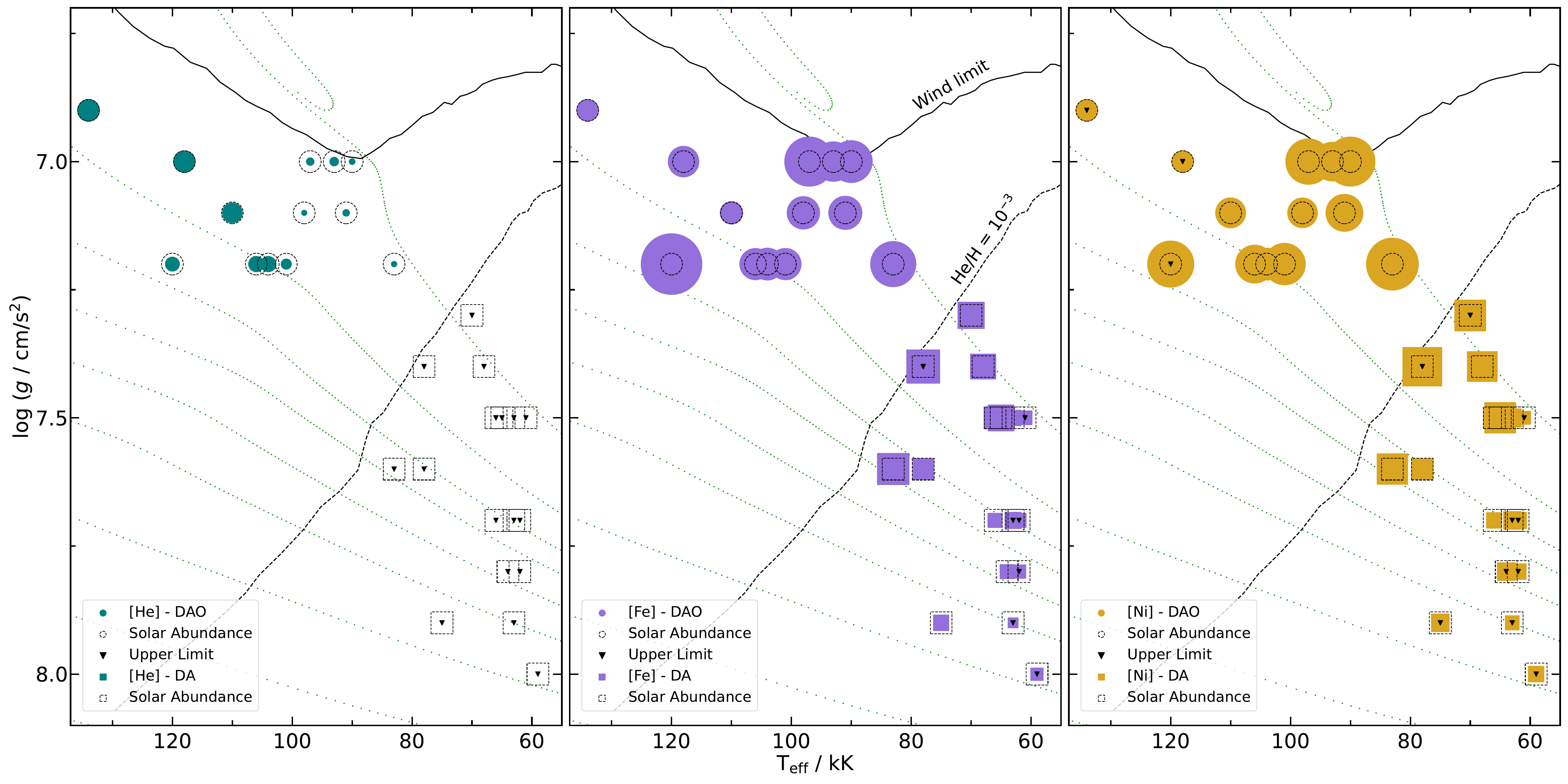}
    \caption{Similar to Fig.~\ref{fig:CNO_Kiel}, but for helium, iron, and nickel (from left to right).}
    \label{fig:HeFeNI_Kiel}
\end{figure*}


\begin{figure*}[h!]
\centering
  \includegraphics[width=17cm]{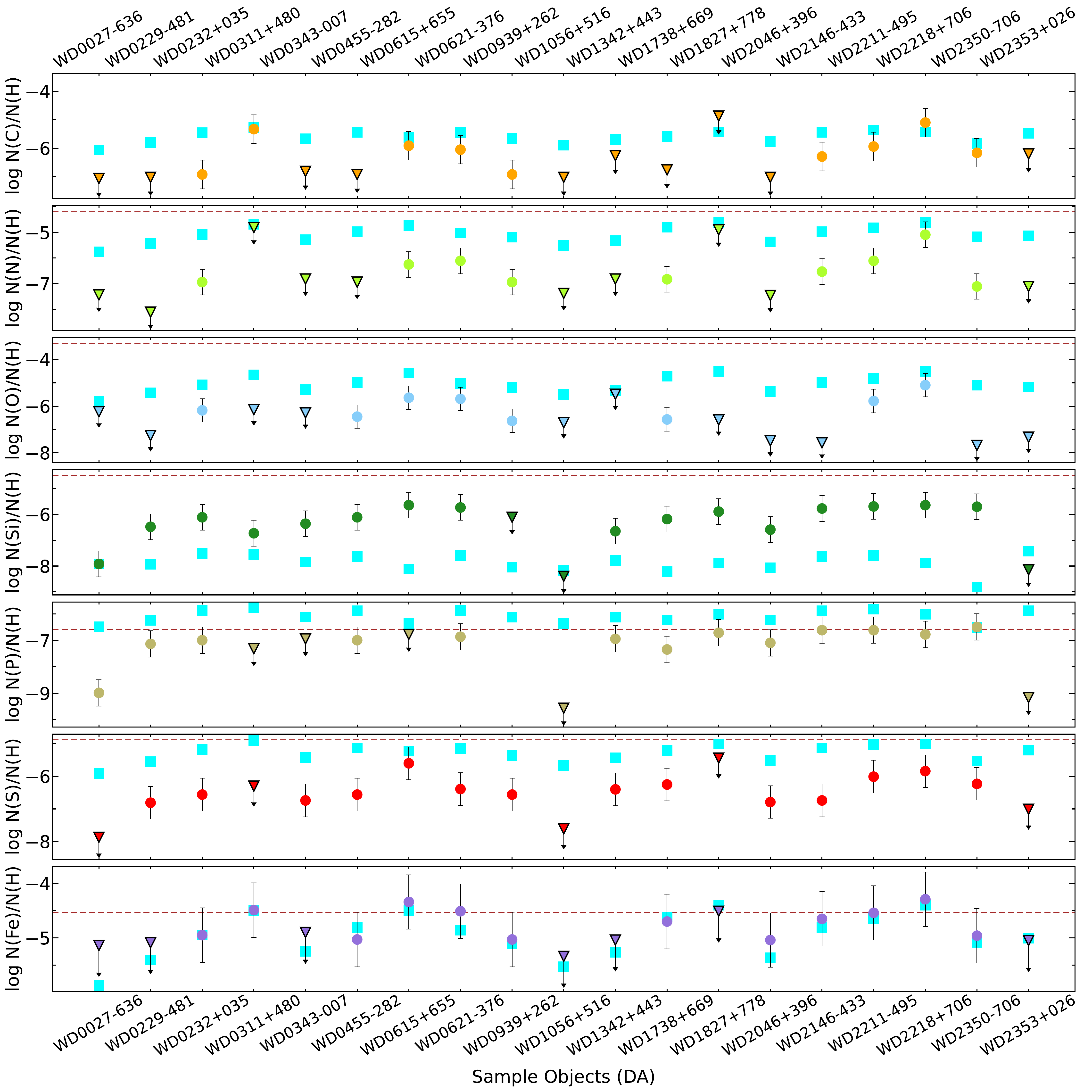}
    \caption{Comparison of the C, N, O, Si, S, and Fe abundances (from top to bottom) in our DAs with predictions of diffusion theory by \citet{1995ApJ...454..429C} and for P by \citet{1996ApJ...468..898V}. The predictions are represented by turquoise squares. The solar abundances are indicated by horizontal dashed lines. The upper limits are marked with triangles.}
    \label{fig:Comp_DA_Chayer}
\end{figure*}

\section{Results and discussion}
\label{sec:discussion} 

In this section, we present the general outcome of the spectral analysis and 
discuss the implications of the results. \teff, \logg, and the mass fractions of the elements are listed in Table \ref{tab:DAO_DA_abundances}. The abundance of a particular element for all objects is also depicted in the Kiel diagram in comparison to the solar values (Fig.~\ref{fig:CNO_Kiel} and \ref{fig:HeFeNI_Kiel}). Moreover, in Fig.~\ref{fig:Comp_DA_Chayer}, a comparison between metal abundances of DA WDs and theoretical diffusion calculations by \citet{1995ApJ...454..429C} and \citet{1996ApJ...468..898V} is shown. 

\subsection{Mass loss and accretion}
\label{subsec:discussion_ML_Acc}

The winds of hot stars mainly emerge due to radiative forces acting on the metals. The metals in turn transfer momentum with Coulomb collisions to H and He, on which the radiative forces are very low \citep{2020A&A...635A.173K}. \citet{1998A&A...338...75U,2000A&A...359.1042U} showed that chemically homogeneous winds could prevent gravitational settling of He in DAO atmospheres. When the WD crosses the wind limit and mass-loss ceases, He can no longer be sustained in the atmosphere, and a quick transition from DAO to DA is then expected.

This is exactly what we observe. The DAO WDs in our sample exhibit a solar He content at high \teff. As the temperature decreases, the He abundance diminishes as well (Fig.~\ref{fig:HeFeNI_Kiel}). While CNO abundances follow the same trend (Fig.~\ref{fig:CNO_Kiel}), other light metals are slightly subsolar. He is more rapidly depleted than CNO, as predicted by \cite{2000A&A...359.1042U}. In contrast, the Fe and Ni abundances of DAOs exceed the solar value (Fig.~\ref{fig:HeFeNI_Kiel}), excluding Longmore 1, WD\,1111+552, and WD\,1957+225, which are comparable to the Sun. Interestingly, the latter three also display solar He abundance. This implies that the effects of a weak stellar wind are prominent, while gravitational settling did not influence the photospheric abundances \citep{2018A&A...616A..73W}. An increase in the abundances of heavy elements already indicates the influence of the radiative levitation \citep{2016A&A...587A..39R}, which can be noted in the case of WD\,2226$-$210 because it, in contrast, has a lower He abundance than other objects with similar \teff. Therefore, we argue that the abundance pattern of the remaining objects signifies an interplay between diffusion and stellar winds \citep{2018A&A...609A.107W,2020MNRAS.492..528L}.

In contrast to DAOs, CNO is extremely depleted in DA atmospheres, whereas the decrease in other light metals is less severe, especially for P. On the other hand, both Fe and Ni abundances are roughly solar. However, a trend is apparent here: With decreasing \teff and increasing \mbox{\logg}, the Fe and Ni abundances diminish (Fig.~\ref{fig:HeFeNI_Kiel}). This implies that atomic diffusion is the main driving mechanism in the DA atmospheres \citep{2019MNRAS.483.5291W}. However, as we showed in Fig.~\ref{fig:Comp_DA_Chayer}, the comparison of the observed abundance pattern of DA WDs to theoretical diffusion calculations of \citet{1995ApJ...454..429C} and \citet{1996ApJ...468..898V} revealed inconsistencies, except for iron. Different results between observations and the diffusion theory were also previously reported in several studies \citep{1995ApJS...99..189C,1995ApJ...454..429C,2003MNRAS.341..870B,2005MNRAS.363..763D,2005MNRAS.364.1082G,2007A&A...466..317W,2013A&A...560A.106R,2014MNRAS.440.1607B,2019MNRAS.487.3470P}. The common ground in these papers was that at least one more mechanism is required to explain the discrepancy and should be included in theoretical calculations in addition to equilibrium theory. Competing ideas were that either weak mass-loss or accretion shapes the observed abundance pattern.

The mismatch in the Si abundance is particularly curious because Si is the only element that was consistently found in all previous studies to be more abundant than predicted by diffusion theory. \citet{2014MNRAS.440.1607B} stated that overabundant phosphorus and silicon resemble the bulk material of terrestrial planets. The authors argued that the observed metal abundances arise due to accretion from short-lived gaseous disks rather than innately containing them. Then, accreted elements are radiatively levitated in the atmosphere, hence possibly marking the source of accretion as tidally disrupted planetary debris. \citet{2019MNRAS.487.3470P} showed that hot DA WDs in their sample exhibited Fe and Si abundances that were higher by an order of magnitude than the theoretical estimates. The authors interpreted the overabundance of metals as an indication of ongoing accretion from external sources. The existence of gaseous and dusty debris disks around WDs is indeed very well established \citep{2003ApJ...584L..91J,2016MNRAS.455.4467M,2020MNRAS.493.2127M}, and there is strong evidence that the debris disks originate in tidally disrupted planetary material \citep{2019Natur.576...61G,2019Sci...364...66M,2022Natur.602..219C}. However, to the best of our knowledge, the \teff of any observed WD hosting a debris disk does not exceed \mbox{$\sim$30 kK} \citep{2014A&A...566A..34K,2020MNRAS.493.2127M,2020ApJ...905...56M,2021MNRAS.504.2707G}.

From the theoretical standpoint, \citet{1997ASSL..214..253C} demonstrated that when either accretion or mass loss is included in the diffusion calculations, the observed overabundance of Si for a WD with \mbox{\teff = 60 kK} and \logg = 7.36 can be explained. Moreover, \citet{1984ApJ...287..868W} explored the possibility of accretion from the M dwarf companion to Feige 24 as the source of the observed element abundances. Assuming a mass-loss rate of 10\textsuperscript{-13} \Mloss for the M dwarf, they calculated the ram pressure in the stellar wind midway and compared this to the radiative pressure gradient of the WD. Accordingly, they demonstrated that the intense radiation field of the WD (\teff $\approx$\,60 kK) can substantially retard the accretion rate and counteract the accretion. Since our sample objects have similar or higher \teff, first, any grain material originating from supposed planetary debris would be sublimated due to the high luminosity of our objects compared to their cooler counterparts \citep{2007ApJ...662..544V}. Second, it is plausible that the intense radiation pressure from the hot WDs in our sample (\mbox{\teff $\geq$ 59 kK}) could halt the infall of the sublimated material, preventing the formation of even short-lived gaseous disks. Finally, it should be considered that the hot DA WD sample of \citet{2014MNRAS.440.1607B} consisted of objects with lower \teff (16--77 kK) than ours, and only $\approx$\,20 \% of them exceed 50 kK. Although the observed Si abundance of a cool Hyades WD (\teff $\approx$\,20 kK) can be explained with radiative levitation, for another object in the same cluster with a similar \teff, accretion from external sources is needed \citep{2014MNRAS.437L..95C}. Therefore, the accretion-impacted abundance pattern of the DAs might only be a viable scenario for cooler objects (at about \teff $\leq$ 30 kK).

In Sect. \ref{subsubsec:Spec_DA_abund_O_stra} we mentioned that it was not possible to fit all oxygen ions simultaneously in the DA WD spectra. When only \ion{O}{VI} resonance lines are fit, the O abundance is far lower, and \ion{O}{IV} and \ion{O}{V} lines cannot be fit properly, and vice versa. It was suggested that inhomogeneous oxygen stratification causes this problem \citep{2000ApJ...544..423V,2001ApJ...553..399V, 2006ASPC..348..209C,2013A&A...560A.106R}. By calculating self-consistent diffusion models, \citet{2013A&A...560A.106R} showed that an improvement can be achieved when the O abundance is only decreased in the outer atmosphere, where the observed weak \ion{O}{vi} resonance lines form. They suggested that a weak mass loss might diminish the O abundance in the outer atmosphere and might cause the observed \ion{O}{vi} resonance line profiles.

However, chemically homogeneous line-driven winds are not expected above \mbox{\logg = 7.0} for a WD with \teff = \mbox{60 kK} and (sub-) solar metallicity \citep{2007ASPC..372..201U}. Fundamentally, abundance patterns of objects with similar parameters should be predicted by the equilibrium between gravitational settling and radiative levitation. Nonetheless, selective winds (or outflows) that only act on metal ions can still be expected \citep{1995A&A...301..823B,2020A&A...635A.173K} and influence the observed metal abundances. \citet{2007ASPC..372..201U} showed that selective winds are predicted for the WDs (\mbox{\logg $\approx$\,7}) with mass-loss rate \Mdot < 10\textsuperscript{-11} \Mloss. Therefore, the discrepancy between observed abundances of DA WDs and theoretical predictions might arise from the exclusion of mass loss in the diffusion calculations \citep{2007ASPC..372..201U,2008A&A...486..923U}.

\subsection{Wind limit}
\label{subsec:discussion_WL}

\cite{2000A&A...359.1042U} commented that if all WD progenitors had the same composition, they should be clearly separated in the Kiel diagram, and the previously observed coexistence of DAO and DA WDs in the same region can be explained by the difference in initial metallicity. A quick glance at Fig.~\ref{fig:Kiel_diagram} reveals otherwise: DAO and DA WDs are clearly separated in the Kiel diagram, and no DA resides above the predicted He abundance limit. Figs.~\ref{fig:CNO_Kiel} and \ref{fig:HeFeNI_Kiel} illustrate the metal abundances of individual objects in the Kiel diagram. In these plots, no substantial deviation in the abundances of DAOs and DAs among each other can be recognized. In comparison, the difference in the abundance pattern from one species to the next is conspicuous. In Fig.~\ref{fig:HeFeNI_Kiel}, the change in the He abundance in the Kiel diagram is depicted. When we accept that solar He abundance indicates a weak mass loss, our observational wind limit falls below the theoretical predictions. Around \mbox{105 kK}, the decrease in He abundance can be observed. This was also reported by \citet{2018A&A...616A..73W,2019MNRAS.483.5291W,2020A&A...642A.228W}. All points above indicate that all hydrogen-rich WDs are born as DAOs and evolve into DAs. Therefore, the initial composition of our sample WDs probably does not substantially deviate in the objects, and the clear detachment of DAOs from DAs supports this claim. However, we should note that the termination point of mass loss probably differs for objects with different metallicities. Therefore, we speculate that finding DAOs far below the He limit, or vice versa for DAs, could be a direct indication of the difference in initial metallicity. 

It is common practice to compile \teff and \logg values from several studies for a broad overview of the evolutionary state of the WDs \citep{1999A&A...350..101N,1998A&A...338...75U,2000A&A...359.1042U}. However, it should be noted that the compiled studies might employ different stellar atmosphere codes and atomic data. Even minor deviations between codes can have a severe impact on the results, as shown by \citet{2008A&A...481..807R}, who reported an (E)UV flux difference between models calculated with TMAP and TLUSTY\footnote{\url{http://nova.astro.umd.edu}} due to different cutoff frequencies of the \ion{H}{i} Lyman bound-free opacity. Therefore, conducting a homogeneous spectral analysis with the same code and atomic data is a key factor in our results. Even though other homogeneously performed analyses identified a tendency for DAOs to be hotter than DAs, a partial overlap of DAs and DAOs in the Kiel diagram persisted \citep{2010ApJ...720..581G,2011ApJ...743..138G,2020ApJ...901...93B}. A result similar to ours can be observed in the \teff -- \logg diagram by \citet{2023A&A...677A..29R}, although several DA WDs inhabit a region slightly beyond the predicted He limit (see their Fig. 6). In all of these cases, the missing ingredient probably is the UV spectroscopy, which enables the inclusion of metal opacities in the model atoms and consequently paves the way to accurately assess \teff by exploiting the ionization balance of metal lines.

\begin{figure}[t!]
\resizebox{\hsize}{!}{\includegraphics[trim=0.5cm 1cm 0cm 14cm, clip]{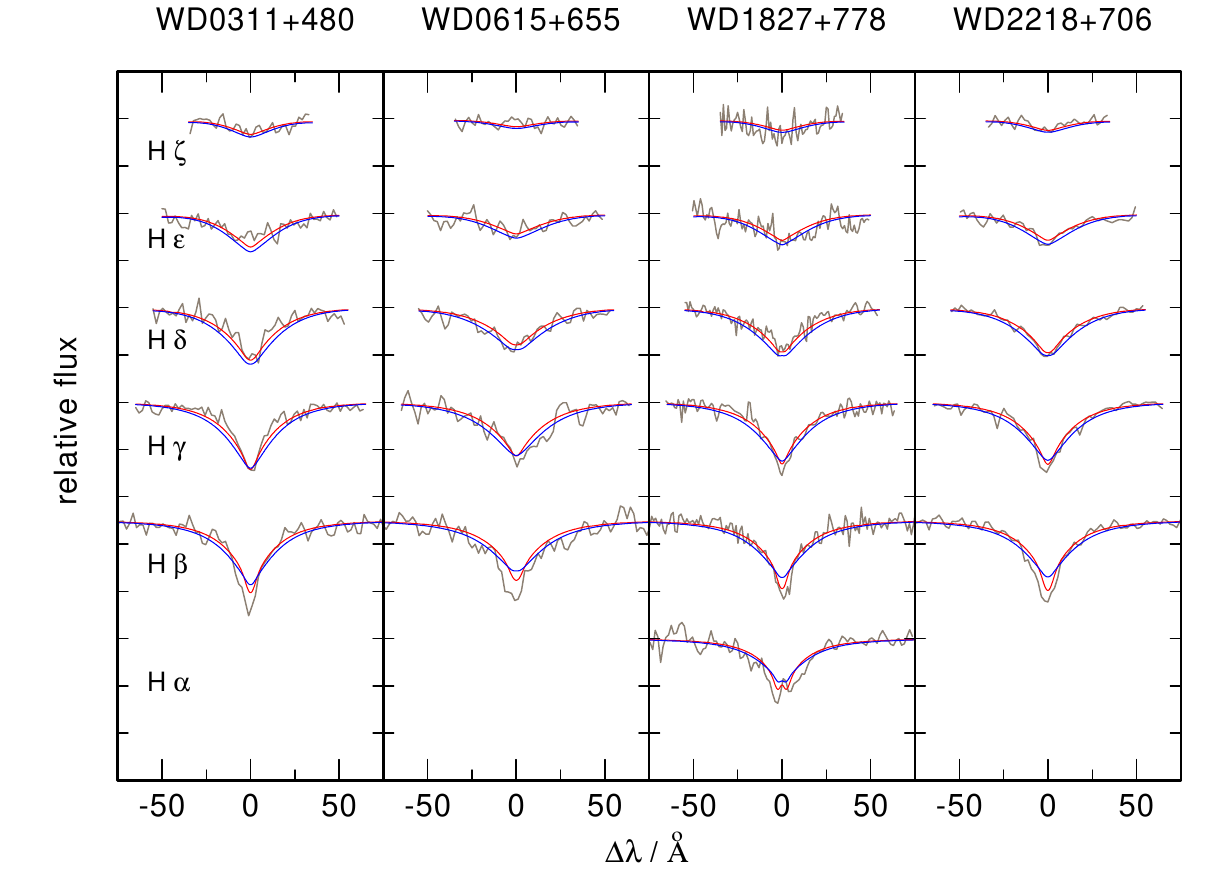}}
  \caption{Balmer lines (gray) of the DA WDs compared to the final model, including metal opacities (red) and pure H models (blue).}
    \label{fig:BLPDA}
\end{figure}

\begin{figure}[t!]
\resizebox{\hsize}{!}{\includegraphics[trim=0.5cm 1cm 0cm 14cm, clip]{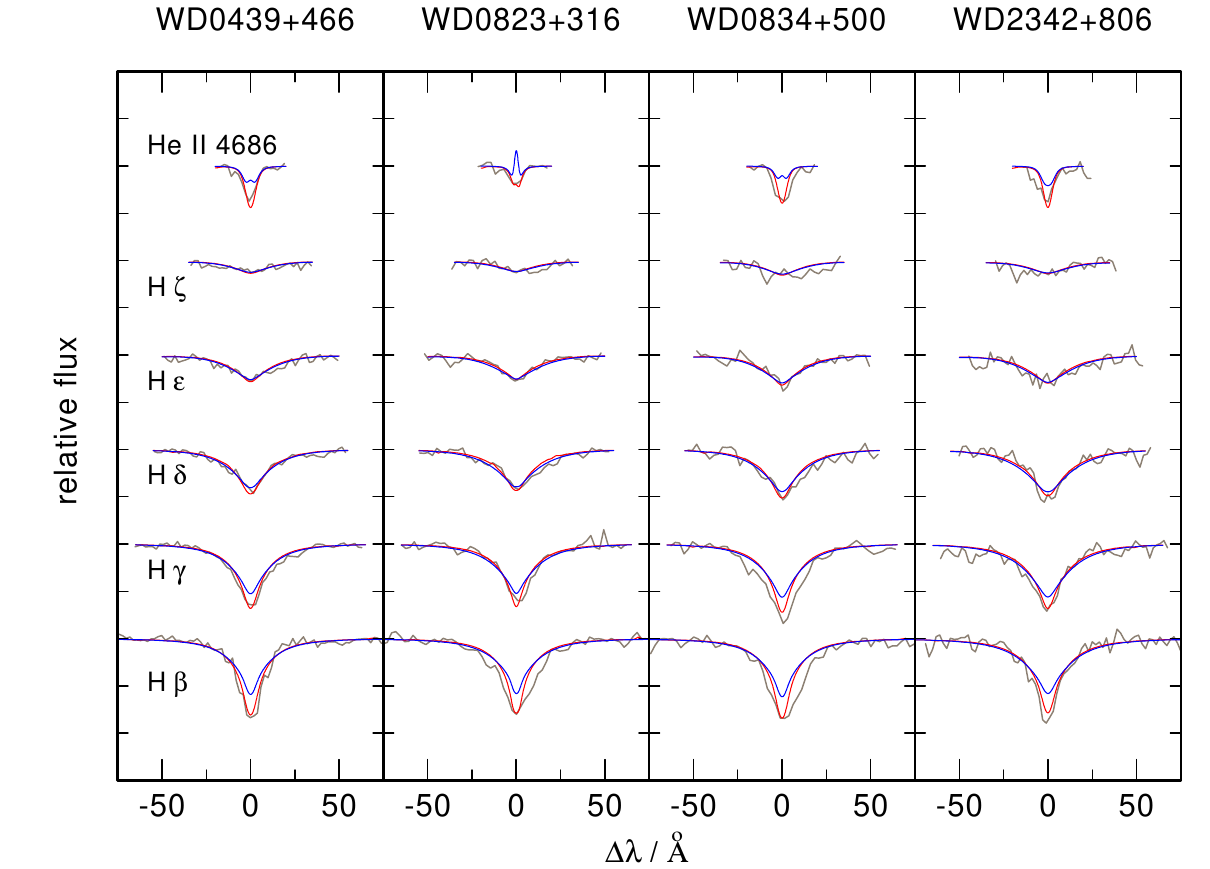}}
  \caption{\ion{He}{II} and Balmer lines (gray) of the DAO WDs compared to the final model, including metal opacities (red) and H+He models (blue).}
    \label{fig:BLPDAO}
\end{figure}

\subsection{Balmer-line problem}
\label{subsec:discussion_BLP}

\begin{figure*}[t!]
\sidecaption
\includegraphics[width=12cm]{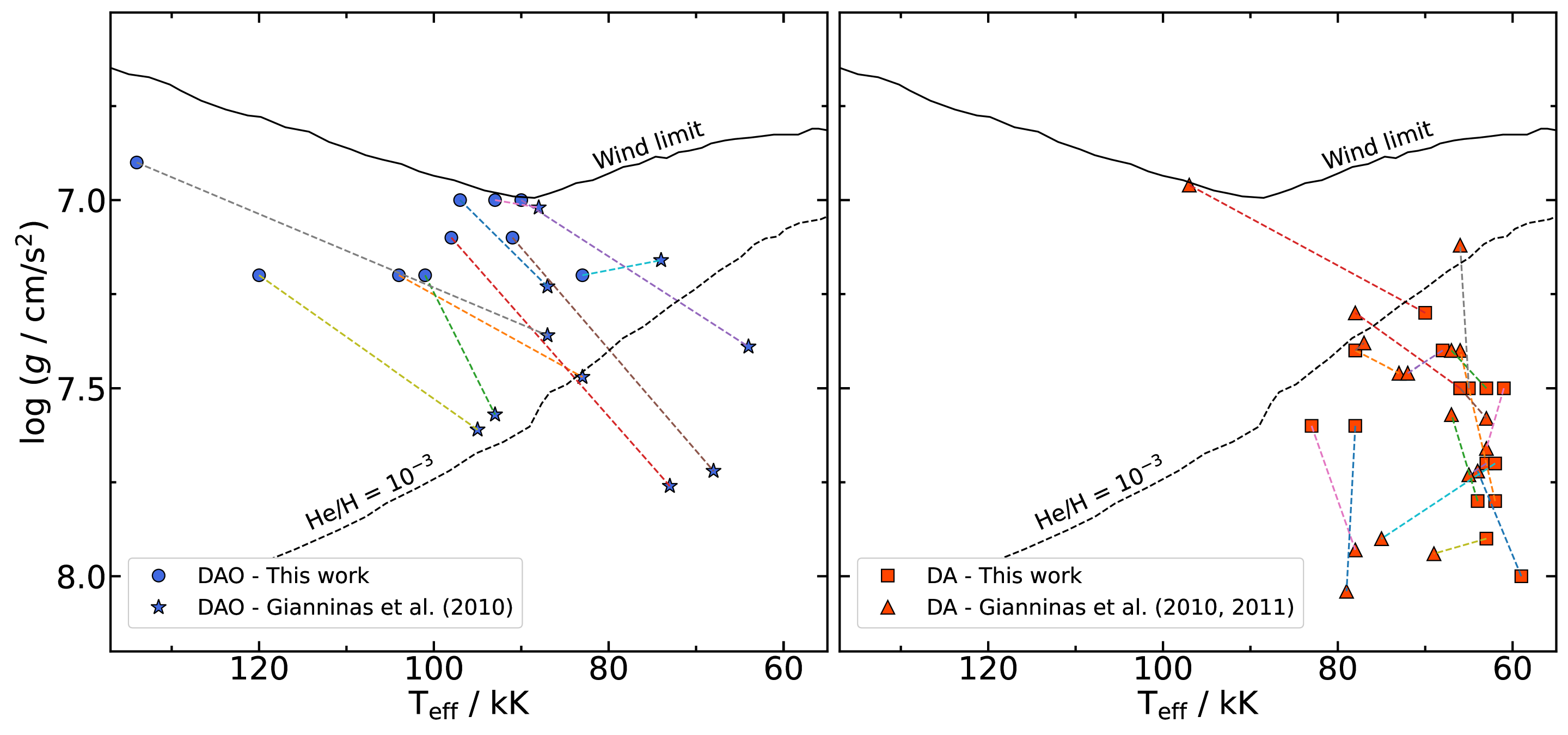}
  \caption{\teff and \logg values of the sample objects (DAO shown as blue circles, and DA shown as red squares) compared to the optical results (DAO shown as blue stars, and DA shown as red triangles) by \citet{2010ApJ...720..581G,2011ApJ...743..138G}.}
    \label{fig:Comp_Giannnias_DAO_DA}
\end{figure*}


The occurrence of the BLP in a large fraction of DAO WDs has already been reported \citep{1994ApJ...432..305B,1999A&A...350..101N,2010ApJ...720..581G, 2020ApJ...901...93B,2023A&A...677A..29R}, and the severity of this problem was tied to the high metal abundance by qualitatively comparing UV spectra of DAO and DA WDs \citep{2010ApJ...720..581G}. To evaluate the impact of the metallicity on the Balmer lines and the BLP, we calculated a set of models with finally adopted parameters that only included H and H+He in the model atoms for DA and DAOs, respectively. With this configuration, we achieved satisfactory fits within the error range for 13 DA WDs. For the remaining 6 DAs, the BLP is more or less apparent with pure H models (Fig.~\ref{fig:BLPDA}). Notably, the latter 6 have a higher metal abundance than the rest of the DA WDs in our sample. Implementing a similar approach for DAOs yielded insufficient fits to \Ionw{He}{2}{4686} and Balmer lines without exception (Fig.~\ref{fig:BLPDAO}). Although we did not try to achieve good fits with this configuration, a tendency for lower \teff can be expected with H+He models. This might explain the large disparity in \teff (up to 40 kK and 20 kK for DAOs and DAs, respectively) and \logg between previous optical studies and ours. However, a large discrepancy in \teff compared to optical results by \citet{2010ApJ...720..581G,2011ApJ...743..138G} is still observed (Fig.~\ref{fig:Comp_Giannnias_DAO_DA}), even though their models include (solar) C, N, and O opacities. While for DAOs, our models systematically predict higher \teff, a general trend for DAs is not apparent.

Considering our models with finally accepted \teff, \logg and meticulously derived metal abundances from UV, we did not encounter severe cases of the BLP in our sample, except for two DAOs (Fig.~\ref{fig:BLPDAO}; WD\,0823+316 and WD\,0834+500) and two DA WDs (Fig.~\ref{fig:BLPDA}; WD\,0311+480 and WD\,0615+655), which respectively showed broader and deeper \mbox{H\,$\beta$} than our models predict. However, the issues observed in the last four cases might depend on several factors, for instance, the data quality, data reduction, and normalization of the spectra. On the other hand, the data quality is of critical importance in detecting the BLP, as already remarked by \citet{2020ApJ...901...93B}  and \cite{2023A&A...677A..29R}. A high fraction of MWDD spectra in our sample have a reasonably high S/N ($\approx$\,50), but most of the observations do not contain the H\,$\alpha$ region. Additionally, the NLTE line-core emissions are not resolved because of the low resolution, which acts as a proxy for \teff. \citet{2023A&A...677A..29R} noted that they encountered the BLP more often in the higher-quality X-shooter data than in those from the Intermediate Dispersion Spectrograph (IDS) at the Isaac Newton Telescope (INT). The same outcome was also reported by \citet{2018A&A...616A..73W} in the case of EGB6, for which the BLP was undetected in the low-resolution MWDD spectra \citep{2010ApJ...720..581G}. In contrast, the BLP still manifests in low-resolution, low S/N spectra even when the model atmosphere contains metal opacities \citep{2019MNRAS.483.5291W}. Therefore, our low-rate detection of BLP does not build a clear case for the solution being the inclusion of metals, but metals rather present an improvement to the results. 

Finally, we would like to comment on the observed differences in \logg inferred from the Lyman and Balmer lines. We achieved satisfactory fits to the Lyman lines of WD\,0621$-$373 and WD\,2211$-$495 (see Fig.~\ref{fig:WD2211-495_FUSE} as an example). However, H\,$\alpha$ and H\,$\beta$ cannot be reproduced properly with the same parameters, and a rather large decrease in \logg (0.3--0.4 dex) is needed, while the higher-order Balmer-line series agree. Both problematic objects have UVES and MWDD spectra, and the same effect can be observed in both datasets.  A similar effect but in a more severe form has already been reported, where large scatter between UV and optical parameters is noticeable \citep{2003MNRAS.341..870B, 2004MNRAS.355.1031G}. For objects above 50 kK, \citet{2005ASPC..334..185V} showed that \teff measured from Lyman lines exceeds up to $\approx$15 \%. However, they demonstrated that when heavy elements are included in the model calculations, a decrease of up to 16 kK in the Lyman-line temperature could be expected. Their sample also included WD\,0621$-$373 and WD\,2211$-$495, and for both objects, they derived \teff  $>$ 70 kK and \mbox{$\approx$\,65 kK} with pure H models from UV and optical spectra, respectively. Both of their Lyman-line temperatures are above our error limit, and their \logg values from UV are lower, but consistent with ours within the uncertainty. Our case demonstrates that with the inclusion of metals and employing ionization equilibria, a common ground for \teff can be found for both wavelength regimes. Since the temperature is fixed, our results from different bands do not substantially deviate from one another in general, as opposed to previous studies. In the case of problematic objects, another factor should perhaps be considered (e.g., data reduction or normalization).

\section{Summary and conclusion}
\label{sec:summary}

We analyzed a sample of 19 DA and 13 DAO WDs, which have \teff $>60$ kK and for which UV spectra are available, using state-of-the-art metal-line blanketed NLTE model atmospheres to extract crucial information about the spectral evolution of hot H-rich WDs. The effective temperatures and gravities were derived accurately by exploiting the ionization balances of the metal lines in the UV spectra. The abundances of helium and metals were measured with high accuracy. Our main results are listed below.

\smallskip

(i) In contrast to earlier studies, we find a clear separation of DAs and DAOs in the Kiel diagram (Fig.~\ref{fig:Kiel_diagram}). Hydrogen-rich WDs are born as DAOs and turn into DAs when they cool to \teff $\approx$\,75--85 kK. At around this temperature, helium becomes depleted by gravitational settling, as predicted by the combined theory of diffusion and mass loss. 

(ii) In agreement with theory, we witness a gradual decrease in the helium abundance when DAOs cool (Fig.~\ref{fig:HeFeNI_Kiel}). The abundances of CNO elements also decrease, but less rapidly, again in accordance with theory (Fig.~\ref{fig:CNO_Kiel}). Other light metals (Si, P, and S) follow a similar trend (Fig.~\ref{fig:CNO_Kiel}).

(iii) According to the He abundance pattern of DAOs, the observational wind limit is slightly shifted downward in the Kiel diagram compared to the theoretical predictions (Fig.~\ref{fig:HeFeNI_Kiel}).

(iv) As DAOs approach the transformation into DAs, an increase in the iron and nickel abundances is observed (Fig.~\ref{fig:HeFeNI_Kiel}). Along with the decrease in the abundance of other elements, this indicates that mass loss can no longer efficiently homogenize the atmosphere.

(v) Diffusion theory assuming equilibrium of radiative acceleration and gravity generally fails to explain the observed metal abundances in hot DAs. This has been stated in many earlier works. However, we report clear systematic trends in our homogeneous sample. For virtually all DAs in our sample, theory overpredicts the abundances of C, N, O, P, and S, but underpredicts the abundance of Si (Fig.~\ref{fig:Comp_DA_Chayer}). Theory and observation agree for iron. The causes of the discrepancies remain unclear, but, as reasoned before by others, residual weak mass loss might be the cause.

(vi) There is no indication that the accretion of circumstellar material is relevant in our sample. All DAs have a similar metal abundance pattern.

(vii) The Balmer-line problem in DAOs is mitigated by the inclusion of metals with abundances determined from the UV analyses, but it does not disappear.

(viii) In contrast to earlier work, we generally encountered no large discrepancies between Balmer- and Lyman-line fits; hence, we find no indication of problems with the line-broadening theory. Larger deviations in gravity determinations are detected for a few objects, but this may be introduced by uncertainties in data reduction and normalization of the optical spectra or by strong ISM line blending in the case of UV spectra.

\smallskip

In conclusion, we emphasize that UV spectroscopy is the key to a reliable parameter determination of hot hydrogen-rich WDs and, hence, their spectral evolution.

\begin{description}[wide,itemindent=\labelsep]
    \footnotesize
    \item[Data availability.]The model fits to the UV and optical spectra of the sample objects as well as the comparison of \teff, \logg, and abundances to previous studies are available at \url{https://doi.org/10.5281/zenodo.13940390} as online supplementary material. 
\end{description}

\begin{acknowledgements}
We thank the referee for a constructive report that helped to improve the paper. We also thank Simon Preval and Martin Barstow for sending us the STIS spectrum of WD0455-282. S.F. is supported by the Deutsche Forschungsgemeinschaft (grant WE1312/58-1). N.R. is supported by the Deutsche Forschungsgemeinschaft (DFG) through grant RE3915/2-1. The TMAD (\url{http://astro.uni-tuebingen.de/~TMAD}) and TIRO tool (\url{http://astro.uni-tuebingen.de/~TIRO}) used for this paper was constructed as part of the activities of the German Astrophysical Virtual Observatory. Some of the data presented in this paper were obtained from the Mikulski Archive for Space Telescopes (MAST). This work is based on observations made with the NASA/ESA \emph{Hubble} Space Telescope obtained from the Space Telescope Science Institute (STScI), which is operated by the Association of Universities for Research in Astronomy, Inc., under NASA contract NAS 5–26555. Funding for the SDSS and SDSS-II has been provided by the Alfred P. Sloan Foundation, the Participating Institutions, the National Science Foundation, the U.S. Department of Energy, the National Aeronautics and Space Administration, the Japanese Monbukagakusho, the Max Planck Society, and the Higher Education Funding Council for England. The SDSS Web Site is \url{http://www.sdss.org/}. This work is based on data obtained from the ESO Science Archive Facility. This research has made use of NASA’s Astrophysics Data System and the SIMBAD database, operated at CDS, Strasbourg, France. This research has made use of the VizieR catalogue access tool, CDS, Strasbourg, France. This research made use of TOPCAT, an interactive graphical viewer and editor for tabular data \citep{2005ASPC..347...29T}.
\end{acknowledgements}

\bibliographystyle{aa} 
\bibliography{bibfile} 


\begin{appendix}

\label{Appendix:Notes}
\section{Remarks on individual objects}
In this section we give a brief remark on each object and compare our spectroscopic results to previous analyses (Tables \ref{tab:compareDAO} and \ref{tab:compareDA}). If the same authors analyzed a particular object more than once, only the latest reported \teff and \logg values were compared, excluding results acquired from a different waveband. 

\subsection{DAO white dwarfs}

\begin{description}[wide,itemindent=\labelsep]
\item[\textbf{Longmore~1}] For the central star of Longmore 1, we found \mbox{\teff = 118 $\pm$ 5 kK} and \loggpm{7.0}{0.3} which are in agreement with the values determined in other FUV analyses \citep{2004PASP..116..391H,2012PhDT.......152Z}. This is the only object in our sample that shows C, N, and O enhancement (up to 1.1 dex). We determined solar He, P, S, Ni, and almost two times solar Fe abundances. All abundance measurements of \citet{2012PhDT.......152Z} lie within our error limits except for N and Fe for which the discrepancy approaches $\approx$\,1\,dex. Likewise, the abundance of several ions (C, N, Si) differs more than 0.5 dex from the results of \citet{2004PASP..116..391H}. Additionally, multiple \ion{C}{iv}, \ion{N}{iv}, \ion{N}{v}, \ion{O}{v}, and \ion{O}{vi} lines were identified in the UVES spectrum. Especially, \ion{O}{V} and \ion{O}{vi} lines in the optical range and NLTE-core emission of \Ionw{He}{2}{4686} and \mbox{H\,$\alpha$} are extremely sensitive to temperature change, and our fit to the UVES spectrum affirms our measured parameters. Nonetheless, we achieved poor fits to  H\,$\beta$, H\,$\gamma$, and \Ionw{He}{2}{5412}. This issue does not appear as a case of conventional Balmer-line problem since we achieve very good fits to \Ionw{He}{2}{4686} and  H\,$\alpha$ as well as the higher order of Balmer-line series. Utilizing evolutionary tracks of \citet{2016A&A...588A..25M} revealed a post-AGB age of 36 $\pm$ 14 kyr for the central star, which is consistent with the expected lifespan of PNe.

\item[\textbf{WD\,0439+466}] is the DAO-type central star of the planetary nebula (PN) Sh 2-216. We estimate \mbox{\teff = 97 $\pm$ 5 kK} and \logg = 7.0 $\pm$ 0.2. \citet{1999A&A...350..101N} reports \mbox{\teff = 83.2 $\pm$ 3.3 kK} and \mbox{\logg = 6.74 $\pm$ 0.19} from their NLTE analysis of optical spectra consisting of H+He models. Another optical study by \citet{2010ApJ...720..581G} reached a similar conclusion for \teff and \logg (\mbox{87 kK, 7.23}). The slight difference to our results can be attributed to the exclusion of metal opacities in the previous optical analyses. On the other hand, our results are in agreement with the values determined by \citet[\mbox{\teff = 93 kK}, \mbox{\logg = 6.9}]{2005ASPC..334..325T} and \citet[\teffpm{95}{2}, \loggpm{6.9}{0.2}]{2007A&A...470..317R}, who employed HST and FUSE spectra. Our study indicates 0.9 dex sub-solar He abundance, whereas C and O are slightly sub-solar, and N is solar. In contrast to light metals, we found Fe and Ni enhancement for this object \mbox{(0.6--0.7 dex)}. \citet{2005ASPC..334..325T} derived abundances only for He, C, N, O, and Si. Among these ions, only N abundance substantially differs from our results. Also, our abundances agree very well with those from \citet{2007A&A...470..317R}, the largest difference being less than 0.4 dex for Si.

\smallskip

\item[\textbf{WD\,0500$-$156}] Optical and UV spectra of the DAO-type central star of PN Abell 7 were investigated multiple times since it has been analyzed by \citet[\mbox{\teff = 75 $\pm$ 10 kK}, \loggpm{7.0}{0.5}]{1981A&A...101..323M}. Earlier optical studies with LTE and NLTE models reported a wide \teff (60--100 kK) and \logg (6.60--7.47) range \citep{1985ApJS...58..379W,1997IAUS..180..120M,1999A&A...350..101N,2004MNRAS.355.1031G,2010ApJ...720..581G}. Although FUV analyses reach a smaller discrepancy in \teff (99--109 kK), the \logg difference (7.00--7.68) remains large \citep{2004MNRAS.355.1031G,2012PhDT.......152Z}. From our FUV plus optical study, we determined \teffpm{104}{6} and \loggpm{7.20}{0.2}. We found the He abundance as half of the solar value. C and O abundances are around the solar value, whereas N is slightly depleted. Other light metals (Si, P, S) were also found as sub-solar, while Fe and Ni are lightly enhanced. In general, our abundances agree very well with other studies \citep{2005MNRAS.363..183G,2012PhDT.......152Z}; however, C and N abundances by \citet{2005MNRAS.363..183G} reside slightly outside of our error limit.

\smallskip

\item[\textbf{WD\,0615+556}] is the DAO-type central star of PN PuWe 1. We derived \teffpm{101}{5} and \loggpm{7.2}{0.2}. In general, \teff difference is less than 10 kK compared to previous optical or UV analyses \citep{1997IAUS..180..120M,1999A&A...350..101N,2010ApJ...720..581G, 2012PhDT.......152Z} and only \logg value of \citet{2010ApJ...720..581G} and \citet[from Lymann lines]{2004MNRAS.355.1031G} lies 0.2 dex out of our error limit. We did not find an agreement with the optical study of \citep{2004MNRAS.355.1031G}, estimating \teffpm{74.2}{4.8}, \loggpm{7.02}{0.20}. He abundance was determined to be one-fifth of the solar value. We found that N, O, Si, and S are slightly depleted, whereas P, Fe, and Ni are 0.3 to 0.5 dex enhanced. An upper limit to C abundance can be assigned as the solar value. Abundance measurements of \citet{2012PhDT.......152Z} are similar to ours, the only statistically significant difference being the P and S abundance. We find a larger disagreement with values from \citet{2005MNRAS.363..183G}, except for Ni abundance.

\smallskip

\item[\textbf{WD\,0823+316}] is the DAO-type central star of PN TK 1. We estimate \teffpm{98}{5} and \loggpm{7.1}{0.2}. The He abundance was derived as \logNfracE{He}{-2.41}, one of the lowest in the sample. We found solar C and P abundances, N is slightly over-solar, and O is slightly sub-solar. Si and S are highly depleted. We encounter over-solar Fe and Ni abundances. Previous optical analyses of this object yielded lower \teff values ranging from 64 kK to 79 kK \citep{1994ApJ...432..305B,2004MNRAS.355.1031G,2010ApJ...720..581G,2011ApJ...730..128T,2020ApJ...901...93B}. However, we found a perfect agreement with the only other UV analysis by \citet[\teffpm{99}{4}, \loggpm{7.26}{0.07}]{2004MNRAS.355.1031G}. Our abundances also assent with \citet{2005MNRAS.363..183G}, except for C and O, which are $\sim$6 and 50 times less than ours.

\smallskip

\item[\textbf{WD\,0834+500}] is the second coolest DAO-type WD in our sample with \teffpm{90}{3}, \loggpm{7.0}{0.2} and \logNfracE{He}{-2.33}. We achieved an excellent fit to the FUSE spectrum, resulting in slightly sub-solar to solar light metal and over-solar Fe and Ni abundances. In contrast to the FUV, our optical fits are not in perfect coherence for given parameters. The line core of H\,$\gamma$ is deeper, and the overall line strength is stronger than our model. On the other hand, our model gives a satisfactory fit to the line core of $\beta$, yet the equivalent width of the line is larger than the model predicts. However, higher orders of the Balmer-line series fit very well, indicating an accurate \logg measurement. Former optical studies indicate much lower \teff, around 60 kK; in contrast, \logg values always reside within our error limits \citep{1994ApJ...432..305B,2004MNRAS.355.1031G,2010ApJ...720..581G}. This is anticipated since not including metal opacities is compensated with a lower \teff.

\smallskip

\item[\textbf{WD\,0851+090}] We derived \teffpm{106}{5}, \logg = 7.2 $\pm$ 0.2 for the DAO-type central star of \mbox{Abell 31}. We encounter the same issue as with Longmore 1 in the UVES spectrum of Abell 31. All of the Balmer and He II line fits are satisfactory except for H\,$\gamma$ and \Ionw{He}{2}{5412}. We estimate the He abundance as half of the solar value. C, N, and O are determined as roughly solar, but Si and S are up to 0.7 dex sub-solar. In comparison, Fe and Ni are slightly enhanced. Our abundances match very well with \citet{2012PhDT.......152Z}, S and Fe being the only ions with a large discrepancy. While \citet{2005MNRAS.363..183G} reported significantly (76 and 20 times) lower N and O content than ours, for other elements, deviations are less than one dex. Additionally, previous studies largely deviate from one another regarding \teff and \logg. On one hand, Balmer fits of \citet{2004MNRAS.355.1031G} resulted in \teffpm{74.7}{6.0}, \loggpm{9.95}{0.15}, on the other hand, \citet{2012PhDT.......152Z} inferred \teffpm{114}{10}, \loggpm{7.4}{0.3} from FUSE spectra.

\smallskip

\item[\textbf{WD\,1111+552}] is the only object for which we did not have optical spectra. We only utilized FUSE observations of the DAO-type central star of PN M 97 (NGC 3587) and determined \teffpm{110}{5} and \loggpm{7.1}{0.3}. Other than Si and S, which are 0.7 and 0.8 dex sub-solar, element abundances are derived as roughly solar. He, N, and O abundances agree well with the PN abundances derived by \citet{2006ApJ...651..898S}, but Fe, which is roughly two dex higher than inferred from PN by \citet{2009ApJ...694.1335D}.  While in general, these measurements, including \teff and \logg, are in good agreement with \citet{2012PhDT.......152Z}, we estimate $\approx$\,12 times more N content. Furthermore, \teff and \logg parameters are comparable to early optical analyses \citep{1997IAUS..180..120M},  with only \citet{1999A&A...350..101N} reporting a temperature 10 kK below our error range. 

\smallskip

\item[\textbf{WD\,1214+267}] is the third coolest DAO WD in our sample (\teffpm{91}{3}, \loggpm{7.1}{0.3}). While early optical studies reported 20 to 30 kK lower temperature and 0.5 to 0.6 dex higher surface gravity \citep{1994ApJ...432..305B,2004MNRAS.355.1031G,2010ApJ...720..581G}, by fitting Lyman lines in the FUSE spectra, \citet{2004MNRAS.355.1031G} conclude \teffpm{87.6}{3.7}, \loggpm{6.96}{0.04}. \citet{2015MNRAS.454.2787G} derived \teffpm{80.3}{4.9}, \loggpm{7.05}{0.20} by utilizing LAMOST spectra; however, they misclassified the object as DA WD. We estimate a roughly one dex sub-solar He abundance. All light metals are sub-solar, with S being the least abundant (one dex sub-solar) among them, whereas heavier elements are slightly enhanced. \citet{2005MNRAS.363..183G} reported only C, O, Si, and Fe abundances. Among them, only O drastically differs from our measurements.

\smallskip

\item[\textbf{WD\,1253+378}] Since it has been discovered \citep{1979A&A....71..163K}, the prototype DAO WD HZ 34 was observed and analyzed multiple times, generating a wide parameter range for temperature (60--91 kK) and surface gravity (6.51--7.02) of the object \citep{1985ApJS...58..379W,1994ApJ...432..305B,1999A&A...350..101N,2004MNRAS.355.1031G,2010ApJ...720..581G}. With  \teffpm{93}{4} and \loggpm{7.0}{0.2}, our results reside at the high end of this parameter range. Additionally, we found 0.7 dex sub-solar He abundance. While we measured slightly sub-solar C, O, and P abundances, N, Si, and S are 10 times sub-solar. On the other hand, $\approx$\,0.5\,dex Fe and Ni enhancement was detected compared to the Sun. Besides He and Si, our abundances significantly differ than \cite{2005MNRAS.363..183G}.

\smallskip

\item[\textbf{WD\,1957+225}] is the hottest DAO-type CSPN (NGC 6853) in our sample (\teffpm{134}{10} and \loggpm{6.9}{0.4}). We found roughly solar abundances for all elements except for Si, which is $\approx$\,0.5\,dex sub-solar. PN abundances (He, N, and O) derived by \citet{2006ApJ...651..898S} reside within our error range. In general, our \teff, \logg and abundance values are consistent with previous FUV analyses \citep{2005ASPC..334..325T,2012PhDT.......152Z}, except for Fe abundance reported by \citet{2012PhDT.......152Z} which is 1.5 dex sub-solar. However, \teff and \logg parameters derived in optical analyses reveal notable differences. For instance, the disparity in \teff, considering our lower limit, extends to $\sim$15 kK compared to \citet[\teffpm{108.6}{6.8}  \loggpm{6.72}{0.23}]{1999A&A...350..101N}. The deviation is even larger in comparison to \citet[\teffpm{86.7}{5.4}, \loggpm{7.36}{0.18}]{2010ApJ...720..581G}. We encountered issues fitting the Balmer lines, especially H\,$\beta$, which is significantly stronger than our models. However, we appoint this to over-correction of the spectrum during background subtraction because forbidden nebular lines (O[III]) are also in absorption instead of emission. This issue was also reported by \citet{2010ApJ...720..581G} and might be the reason for their significantly lower \teff.

\smallskip

\begin{figure*}[h!]
\centering
  \includegraphics[width=18cm,trim=0cm 7.5cm 0cm 1.5cm, clip]{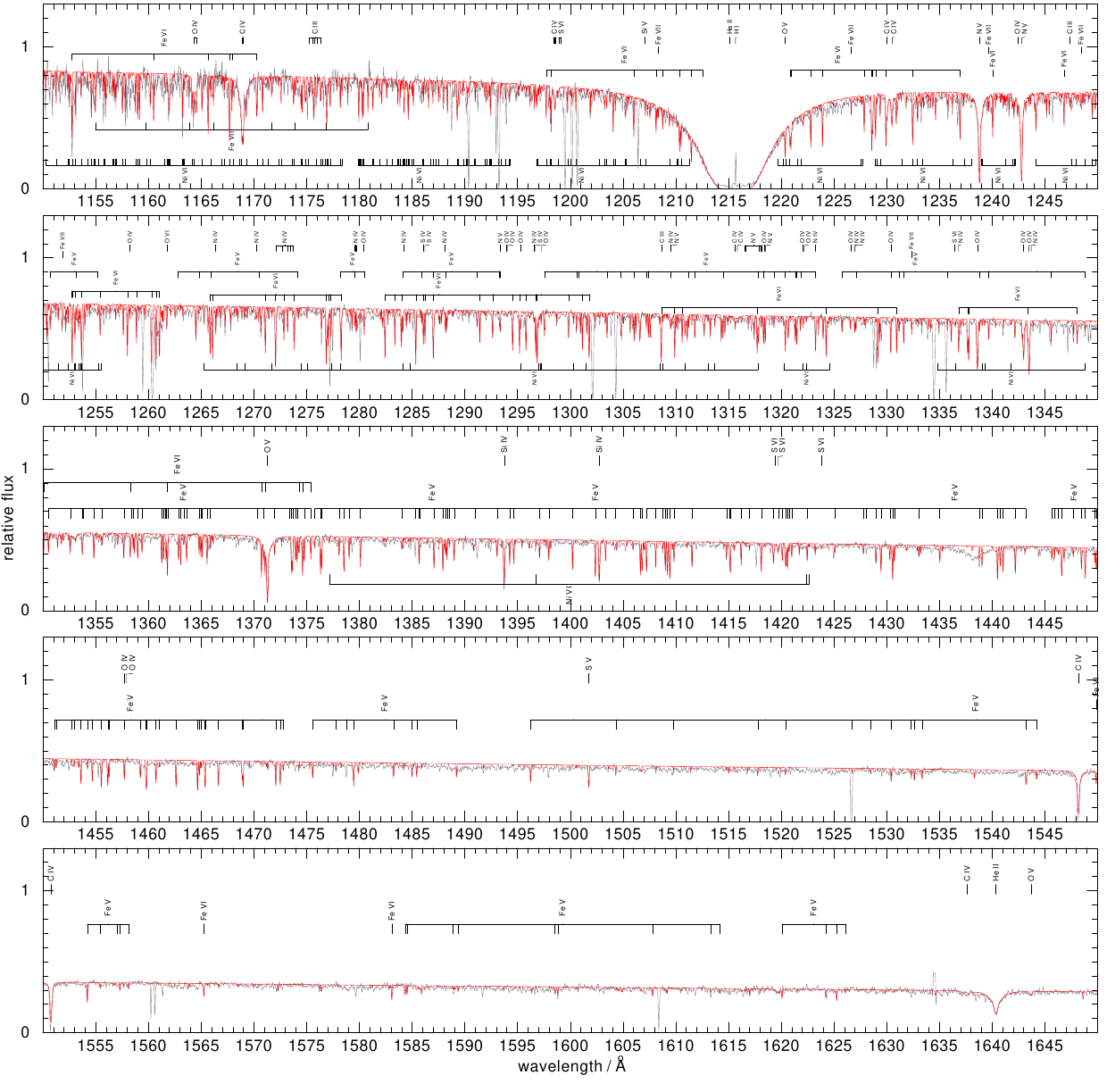}
    \caption{HST spectrum of the DAO WD\,2342+806 (red: model with \teff = 83 kK, \logg = 7.2).}
    \label{fig:WD2342+806_HST}
\end{figure*}

\item[\textbf{WD\,2226$-$210}] is the central star of Helix nebula (NGC 7293). We derived \teffpm{120}{5} and \loggpm{7.2}{0.3} and 0.4 dex sub-solar He abundance. O, Si, and S are slightly depleted, and other light metals are comparable to solar abundance. On the other hand, Fe and Ni are enriched by 0.9 and 0.7 dex super-solar, respectively. S abundance determined from PN by \citet{2004AJ....127.2284H} is in agreement with the photospheric abundances, whereas He, N, and O abundances lie 0.1 dex above our error limit. Aside from the substantial \logg difference (0.9 dex) with \citet{2005ASPC..334..325T}, all parameters from other FUV analyses \citep{2005ASPC..334..325T,2012PhDT.......152Z} remain within our error range. While previous optical studies report lower \teff in a consistent manner, ranging from 90 kk to 104 kK \citep{1985A&A...142..289M,1988A&A...190..113M,1997IAUS..180..120M,1999A&A...350..101N,2010ApJ...720..581G}, \logg values differ significantly (6.6--7.6 dex).
\smallskip

\item[\textbf{WD\,2342+806}] is the coolest (\teffpm{83}{5}, \logg = 7.2 $\pm$ 0.2) DAO-type WD (GD 561) with one of the lowest He content (1.2 dex sub-solar) in our sample. C, N, O, Si, and S abundances range between  $\sim$ 0.2 to 0.5 dex sub-solar, whereas Fe and Ni are enriched up to 6 times the solar content. He, Si, P, and S abundances determined by \citet{2012PhDT.......152Z} remain out of our error limits; other species agree. Except for Fe, our results are consistent with \citet{2005MNRAS.363..183G}. All previous studies report lower \teff and \logg values than ours. 

GD 561 was associated with PN Sh 2-174 \citep{1993IAUS..155..495N,1994AJ....108..978T}. \citet{1994ApJ...432..305B} suggests that the spectroscopic parameters are consistent with a post-EHB star that did not go through the AGB phase. \citet{1999A&A...350..101N} argues that any ejected material on the red giant branch (RGB) would not persist due to long evolutionary time scales on the horizontal branch, and the mass of the remaining EHB star would be too low to create the associated nebula. \citet{1994AJ....108..978T} reports the possibility of GD 561 being a low-mass post-RGB object that was produced through binary interaction, which, in that case, makes Sh 2-174 the envelope lost in the common envelope evolution. However, the vast disparity between the kinematical age of the PN and the supposed post-RGB lifetime of the WD \citep{1999A&A...350..101N} decreases the likelihood. Additionally, the common envelope evolution scenario requires presence of a companion. However, \citet{2005MNRAS.364.1082G} did not detect a statistically significant RV shift. Also, there is no infrared excess, indicating a close companion (Filiz et al. in prep.). \citet{2015ApJ...799..198R} identifies the system as a PN-ISM interaction through their radio analysis. On the other hand, \citet{2006IAUS..234..455M} reports the kinematics of the gas inferred from optical emission lines. \citet{2008PhDT.......109F} concluded that the determined nebular velocity is statistically different from the stellar velocity, and the presence of a strong bowshock is required if the associated star is co-moving with the PN. Instead, they suggested that Sh 2-174 is an HII region in the ambient ISM, mimicking a true PN \citep{2010PASA...27..129F,2016MNRAS.455.1459F}. 

Our estimated mass from the Kiel diagram (0.53 $\pm$ 0.04 \Msol) is higher than previous analyses and indicates that GD 561 is likely a post-AGB star. From the evolutionary tracks of \citet{2016A&A...588A..25M}, we determined a post-AGB age of 230 $\pm$ 73 kyr. Our estimated age of the WD is higher than the expected lifetime (10--100 kyr) of a bona fide PN. This agrees with \mbox{Sh 2-174} being an HII region.

\end{description}

\subsection{DA white dwarfs} 

\begin{description}[wide,itemindent=\labelsep]

\item[\textbf{WD\,0027$-$636}] is the coolest DA WD in our sample (\teffpm{59}{5}, \loggpm{8.0}{0.2}). The given parameters are in unison with former studies, and only \logg values determined by \citet{1997ApJ...480..714V}, \citet{2011ApJ...743..138G} and \citet{2021AJ....162..188B} remain 0.2 dex out of our error range. All light metals are extremely sub-solar, whereas Fe and Ni are lightly depleted. Except for Si and P, we report upper-limit abundances. Only C, O, and Fe values are in agreement with the diffusion calculations of \citet{1995ApJ...454..429C}. To the best of our knowledge, no other abundance measurements have been reported for this object. In their FUV analysis, \citet{2014MNRAS.440.1607B} only report \teff and \logg values. 
\smallskip

\item[\textbf{WD\,0229$-$481}] We estimate \teffpm{62}{5} and \logg = 7.8 $\pm$ 0.2 for this DA WD. With CNO being highly depleted and given as upper limits, the abundance pattern shows a sub-solar content, though the P, Fe, and Ni deficit is about 0.4 dex. Only Fe and P abundances are comparable to theoretical calculations of \citet{1995ApJ...454..429C} and \citet{1996ApJ...468..898V}, respectively. \citet{2014MNRAS.440.1607B} determined the C, Si, P, and S abundances of this object, and except for C, abundances lie within our error limits. Other than \teff reported by \citet{1995ApJ...443..735B} and \citet{2007ApJ...667.1126L}, the highest temperature difference is 5 kK in comparison to previous studies. However, we find larger discrepancies in surface gravity. 
\smallskip
 
\item[\textbf{WD\,0232+035}] Feige 24 is one of the most often analyzed hot-DA WDs with observations from EUV to infrared, of which binary nature has been long known \citep{1966VA......8...63G}. Contamination by the irradiated M dwarf can be recognized in the optical spectra \citep{2011ApJ...743..138G,2015MNRAS.454.2787G}. However, the optical spectrum we obtained from MWDD \citep{2011ApJ...743..138G} does not cover the H\,$\alpha$ region and is not contaminated. As a result of stringent error margins imposed in this work (\teffpm{63}{3} and \loggpm{7.5}{0.2}), only 5 of 15 previously reported \teff and \logg combinations statistically agree with ours, though deviations in \teff do not exceed 8 kK. In contrast, a comparison of abundances with former studies \citep{1992ApJ...392L..27V,2000ApJ...544..423V,2003MNRAS.341..870B,2014MNRAS.440.1607B} reveals a consistent outcome, except for the mismatch in Ni and C abundances derived by \citet{2003MNRAS.341..870B,2014MNRAS.440.1607B}, respectively. CNO being on the lower end, we measured below solar abundances for all elements. Among these, only the Fe abundance is in agreement with the diffusion theory \citep{1995ApJ...454..429C}. Since this object has comparable metal abundances to DA WDs with similar parameters, accretion of wind material from the M dwarf is unlikely.
\smallskip

\smallskip
 
\item[\textbf{WD\,0311+480}] FUSE spectra of this DA WD are heavily contaminated by interstellar absorption, which substantially hinders the measurement of stellar parameters. Only the \ion{C}{iii} and \ion{C}{iv} ionization balance provided a constraint for the temperature. Therefore, setting a large error range was needed (\teffpm{70}{10}). Also, successfully fitting all the Balmer lines was not possible. We managed to replicate line cores other than H $\beta$, which is too deep in the observation. However, line wings of H $\delta$ and H $\epsilon$ are slightly broader in our model. The higher order members of the Balmer-line series, as well as the Lyman lines in the FUSE spectrum, indicate a higher gravity; thus, we opted for \loggpm{7.3}{0.4}. To the best of our knowledge, spectroscopic parameters of this object were previously derived only by \citet[\mbox{\teff = 97.8 kK} and \logg = 6.96]{2010ApJ...720..581G}. Although they utilized three different grids containing H/He, solar-CNO, and 10 $\times$ solar-CNO models, the inclusion of light metals, even in large quantities, did not prevent the Balmer-line problem. Their best-fit parameters were derived from the models that include solar-CNO. However, analyzing FUSE spectra uncovered a sub-solar light metal content, including CNO. In contrast, a slight Fe and Ni enhancement was detected. Except for C, Si, and Fe, we present upper-limit abundances. Only C, N, and Fe abundances yield results similar to predictions of the diffusion theory \citep{1995ApJ...454..429C}. 
 \smallskip

\item[\textbf{WD\,0343$-$007}] We derived \teffpm{63}{4} and \logg = 7.7 $\pm$ 0.2, which are statistically in perfect agreement with previous results from optical spectroscopy \citep{1995A&A...301..823B,1997ApJ...488..375F,2007ApJ...667.1126L,2009A&A...505..441K,2010ApJ...714.1037L,2011ApJ...743..138G}. All of the element abundances were derived as sub-solar, while the depletion is less than 0.3 dex sub-solar for P, Fe and Ni. Excluding Si and S, element abundances are given as upper limits. Only the Fe abundance agrees with the theoretical calculations of \citet{1995ApJ...454..429C}. 
\smallskip
 
\item[\textbf{WD\,0455$-$282}] Previously, EUV, FUV, and optical analyses of this DA WD have been made multiple times. However, we found that only values derived by \citet{2010ApJ...714.1037L} reside within our \teffpm{66}{3} and \loggpm{7.5}{0.2} boundaries in the Kiel diagram. Nonetheless, the temperature differences to our lower and upper limits are smaller than 10 kK and 3 kK, respectively, while early studies except for a few reported a tendency of \mbox{\logg $\geq$ 7.80}. As previously stated, the Balmer-line problem was not detected in our optical fits. We measured sub-solar element abundances, with C, N, O, and Fe, Ni being extremely and slightly depleted, respectively. Only the Fe abundance is consistent with diffusion theory. Our element abundances generally agree with \citet{1996ApJ...468..898V} and \citet{2019MNRAS.487.3470P} except for the differences in P and C, respectively. Also, among C, N, O, Si, Fe, Ni, O and Si are the only elements that show a large discrepancy compared to \citet{2003MNRAS.341..870B}. In contrast, C and S measurements of \citet{2014MNRAS.440.1607B} significantly differ from ours, whereas Si and P\footnote{\citet{2014MNRAS.440.1607B} provide P abundances determined by fitting \ion{P}{iv} and \ion{P}{v} lines. Their measurements from \ion{P}{v} lines are very similar to ours, though abundances determined from \ion{P}{iv} lie $\sim$0.3 dex below our error range.} remain in our error range.
\smallskip
 
\item[\textbf{WD\,0615+655}] is the hottest DA WD in our sample (\teffpm{83}{10} and \loggpm{7.6}{0.4}). While \teff and \logg derived by \citet{1998A&A...338..563H} and \citet{2007ApJ...667.1126L} lie out of our range, our results are in agreement with \citet{2010ApJ...720..581G}. However, we were unable to fit the line core of H\,$\beta$, whereas other Balmer lines are reproduced better. Element abundances were measured by utilizing HST spectra, except for P, which shows no lines in the HST wavelength range. A slightly sub-solar upper limit was estimated for P from FUSE spectra. Fe and Ni are the only species derived above the solar value. The remaining elements were found to be sub-solar. C, P, S, and Fe abundances match with theoretical predictions \citep{1995ApJ...454..429C}.
\smallskip

\item[\textbf{WD\,0621$-$376}] Exploiting multiple ionization balances e.g. \ion{C}{iii}/\ion{C}{iv}, \ion{O}{iv}/\ion{O}{v}, \ion{P}{iv}/\ion{P}{v}, \ion{Fe}{v}/\ion{Fe}{vi}, and \ion{Ni}{v}/\ion{Ni}{vi} enabled us to constrain the temperature tightly (\teffpm{65}{3}). Thanks to high S/N FUSE spectra and dearth of ISM contribution, a perfect fit to Lyman lines could be achieved with \logg=7.5. On the other hand, except for the higher orders, the same parameters resulted in poor Balmer-line fits, emerging as deeper line cores and narrower wings of H\,$\delta$ - H\,$\alpha$. Decreasing \logg to 7.10 solved the issue for deep lines but diminished the quality of higher-order line fits; therefore, imposing a large error limit was inevitable \loggpm{7.5}{0.4}. In comparison, other studies in various bands report a lower \logg, excluding \citet{2003MNRAS.341..870B} and \citet{2019MNRAS.487.3470P} which report similar results. On the other hand, previous abundance measurements reside either within or slightly out of our error limits \citep{1993ApJ...416..806H,2003MNRAS.341..870B,2014MNRAS.440.1607B,2019MNRAS.487.3470P}. However, C and N abundances derived by \citet{2014MNRAS.440.1607B} and \citet{1993ApJ...416..806H} significantly differ from our parameters, respectively. Our study indicates sub-solar light metal and roughly solar Fe and Ni content. C, O, and Fe abundances are comparable to the diffusion theory. 
\smallskip

\item[\textbf{WD\,0939+262}] is the only DA WD in our sample without a FUSE spectrum. Consequently, we could not derive the P abundance for this object since relevant P lines are only found in the FUSE wavelength range. We derived \teffpm{66}{3} and \loggpm{7.7}{0.2}. With Si indicating an upper limit, light metals are heavily depleted compared to the Sun, while Fe and Ni were measured as slightly sub-solar. Only the Fe abundance coincides with the theoretical predictions. All abundances derived by \citet{2003MNRAS.341..870B} are within our error range, but Fe and Ni. Only \teff and \logg determined by \citet{2005ApJS..156...47L} and \citet{2007ApJ...667.1126L} are statistically in agreement within set limits in the Kiel diagram. 
\smallskip

\begin{figure*}[t!]
\centering
  \includegraphics[width=18cm,trim=0cm 11.5cm 0cm 1.5cm, clip]{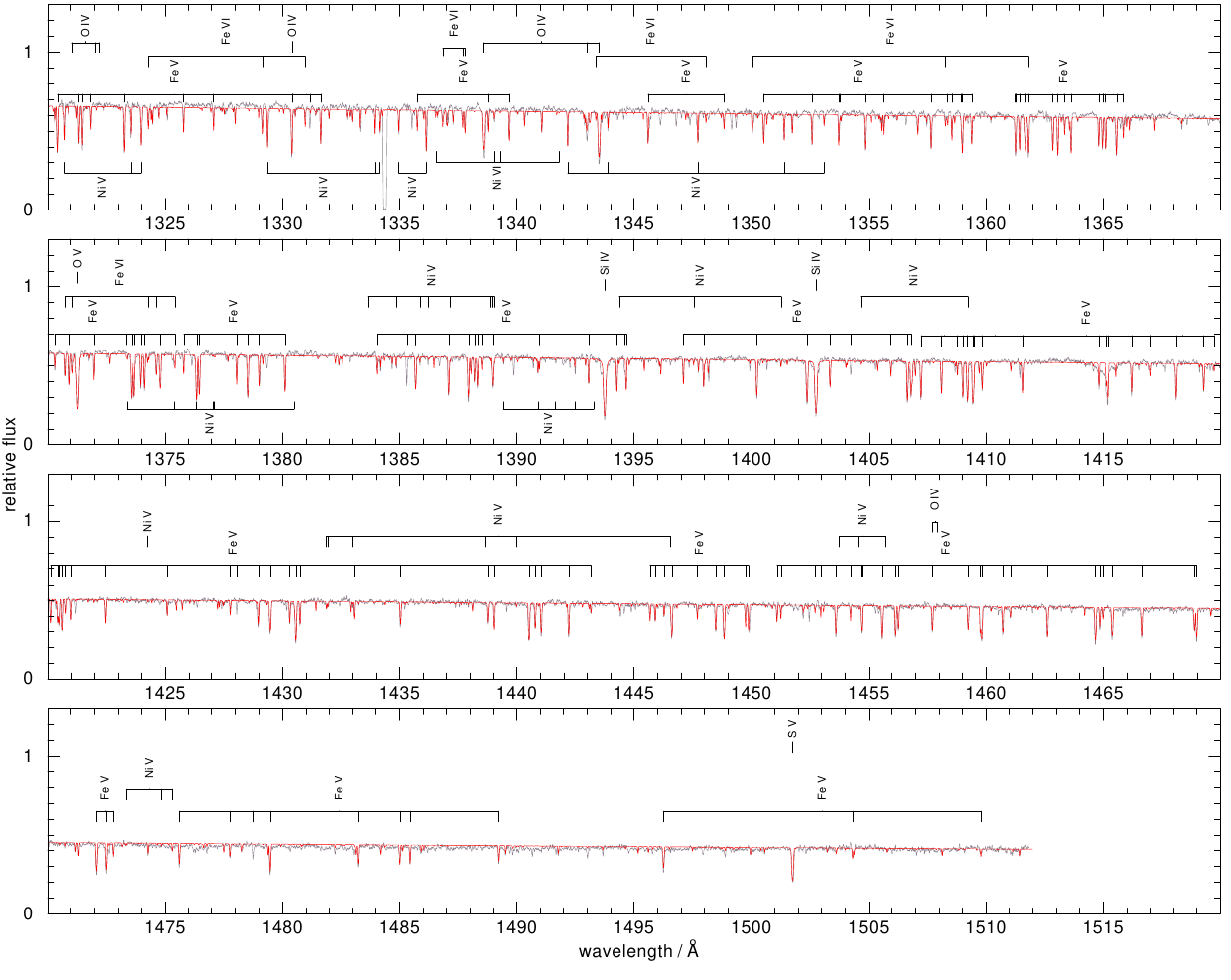}
    \caption{HST spectrum of the DA WD\,2211$-$495 (red: model with \teff = \mbox{68 kK}, \logg = 7.4).}
    \label{fig:WD2211-495_HST}
\end{figure*}

\item[\textbf{WD\,1056+516}] FUSE spectra of this DA WD indicate extremely low metal content. Due to poor data quality, detecting any photospheric line was not possible. Therefore, we increased the element abundances to upper limits where the model lines disappear in the noise of the observation. \citet{2012A&A...546A...1A} reports similar upper limits from the analysis of the FUSE spectra. However, they opt for a lower temperature (\teffpm{56}{2},\loggpm{7.90}{0.3}) due to the EUV flux of the \emph{Chandra} spectrum, which is slightly below our error limit. We estimate \teffpm{63}{5} and \loggpm{7.9}{0.2}. Parameters derived by \citet{1997ApJ...480..714V}, \citet{2007ApJ...667.1126L}, and \citet{2020ApJ...901...93B} through Balmer-line fitting statistically agree with ours. 
\smallskip

\item[\textbf{WD\,1342+443}] We derived \teffpm{62}{5} and \logg = 7.7 $\pm$ 0.3, which statistically only agree with parameters measured by \citet{2014MNRAS.440.1607B}. Both the MWDD and SDSS spectra of this object show unusually weak H\,$\beta$, while we could reproduce the remaining Balmer lines accurately. This might explain the large disparity between FUV and optical analyses, which tend to find higher \teff and \logg. We measured sub-solar abundances for all elements, among which C, N, O, Fe, and Ni were given as upper limits. Our results partially agree with abundances derived by \citet{2003MNRAS.341..870B,2014MNRAS.440.1607B}, who derived C, O, Si, Fe, and Ni; and C, Si, P, and S, respectively. From their first study, O and Si, and their second study, Si, P, and S abundances are comparable to ours. 
\smallskip
 
\item[\textbf{WD\,1738+669}] is one of the hottest DA WDs in our sample (\teffpm{78}{6} and \loggpm{7.6}{0.4}). Temperature and surface gravity derived by \citet{2007ApJ...667.1126L} and \citet{2010ApJ...720..581G} through Balmer-line fitting and by \citet{2003MNRAS.344..562B} through Lyman-line fitting statistically agree with our results. Light-metal abundances were measured as sub-solar, whereas Fe and Ni show solar abundance. Only the Fe abundance is comparable to theoretical predictions. We find an agreement with light metal abundances determined by \citet{2003MNRAS.341..870B}, whereas their Fe and Ni abundances are significantly lower than ours.
\smallskip
 
\item[\textbf{WD\,1827+778}] This object shows a high metal content compared to the rest of the DAs in our sample. All light metals were measured close to solar value except O and Si, whereas Fe and Ni are slightly over solar. However, these results might be deemed questionable due to the poor quality of the FUSE spectra, which also complicated the ionization balance assessment. Therefore, a wider error margin had to be imposed on temperature (\teffpm{78}{10}). Uncertainty on the surface gravity is also large due to the line core of H\,$\beta$ being weaker in our model (\loggpm{7.4}{0.4}). 
\smallskip

\item[\textbf{WD\,2046+396}] We estimate \teffpm{64}{5} and \logg = 7.8 $\pm$ 0.3 for this DA WD. The uncertainty in \logg arises from poor fits to H\,$\alpha$ and H\,$\beta$ that indicate a lower \logg. However, this contradicts what was inferred from Lyman and higher-order Balmer lines. In general, these results are in agreement with previous works \citep{1997ApJ...488..375F,2007ApJ...667.1126L,2011ApJ...743..138G,2011AJ....141...96L}. Additionally, all element abundances were measured as sub-solar, and only the Fe abundance was comparable to diffusion theory. 
\smallskip
 
\item[\textbf{WD\,2146$-$433}] Due to good quality FUSE spectra, we were able to use multiple ionization balances. Hence, we could impose a tight constraint on the temperature (\teffpm{66}{4}). Moreover, metal abundances for this object were measured as sub-solar, excluding P, Fe, and Ni, which are roughly solar. These abundances are in agreement with those derived by \citet{2014MNRAS.440.1607B} except for Si. In addition, Fe and P abundances match with theoretical predictions by \citet{1995ApJ...454..429C} and \citet{1996ApJ...468..898V}, respectively. Generally, both Lyman and Balmer lines indicate a similar gravity (\loggpm{7.5}{0.3}), though a slight mismatch between the observed and model spectrum of H\,$\beta$ enforces a rather high error range for \logg.  
\smallskip

\item[\textbf{WD\,2211$-$495}] We utilized FUSE, HST, UVES, and MWDD spectra to determine the parameters of this object (\teffpm{68}{4} and \loggpm{7.4}{0.3}). Thanks to the high S/N of the UV spectra, \teff and metal abundances could be measured precisely. However, like WD\,0621$-$376, the surface gravity inferred from Lyman and Balmer lines indicates different values. While \mbox{\logg = 7.10} reproduces H\,$\alpha$ and H\,$\beta$ in both MWDD and UVES spectra, \logg = 7.40 perfectly fits Lyman lines and up to H\,$\gamma$. Most of the previous analyses report \teff and \logg within our error limits. Except for P, Fe, and Ni, sub-solar metal abundances were measured. Our abundances are in agreement with those derived by \citet{1993ApJ...416..806H}, \citet{2003MNRAS.341..870B}, and \citet{2019MNRAS.487.3470P}, where only mismatch occurs in N abundance for the former two analyses. On the other hand, excluding S, \citet{2014MNRAS.440.1607B} reports much lower light metal content than ours. Our C and Fe abundances are comparable to theoretical predictions.
\smallskip

\item[\textbf{WD\,2218$+$706}] is the only DA WD in our sample that displays a weak \Ionw{He}{2}{1640} feature. We derived \textit{N}(He)/\textit{N}(H) = 1.3 $\times$ 10$\textsuperscript{-4}$, which is the highest among DAs in our sample and clearly lower than the estimated He abundance limit (Fig.~\ref{fig:WD2218+706_HeII_1640}). We estimate \teffpm{78}{5} and \loggpm{7.4}{0.3}, which statistically agrees only with values derived by \citet{2010ApJ...720..581G}. Moreover, we derived light metal abundances as sub-solar, while Fe and Ni are slightly over solar. Interestingly, WD\,2218+706 is the only DA WD in our sample that shows element abundances close to the theoretical predictions, where C, N, O, and Fe abundances agree very well, and S and P slightly lie outside of our error limits. 
\smallskip

\begin{figure}[t!]
\resizebox{\hsize}{!}{\includegraphics[trim=2cm 1cm 4cm 20.4cm]{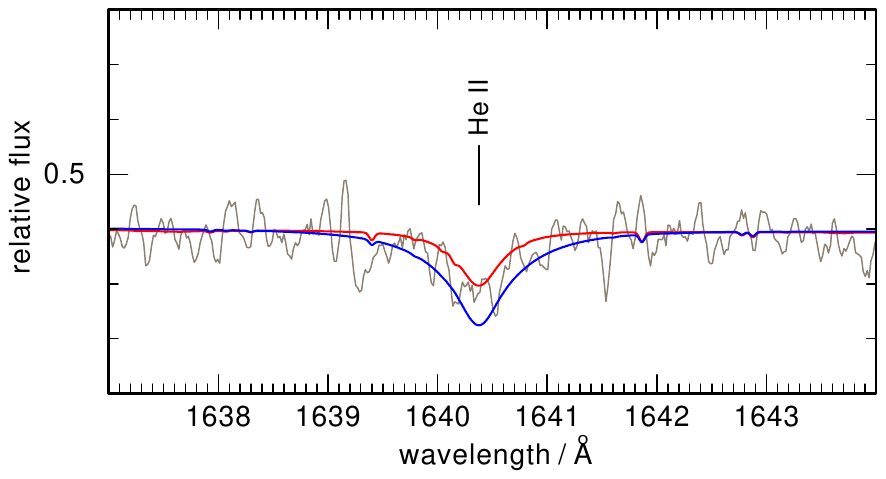}}
  \caption{Section of the STIS spectrum (gray) of the DA WD\,2218+706 compared to models (\mbox{\teff = 78 kK}, \logg = 7.4) with measured He abundance (red; \textit{N}(He)/\textit{N}(H) = 1.3 $\times$ 10$\textsuperscript{-4}$) and theoretical He limit (blue; \mbox{\textit{N}(He)/\textit{N}(H) = 10$\textsuperscript{-3}$}).}
    \label{fig:WD2218+706_HeII_1640}
\end{figure}

\item[\textbf{WD\,2350$-$706}] HD 223816 is a wide binary system consisting of an F5IV star and a DA WD \citep{1994MNRAS.270..499B}, which has been resolved in HST imaging \citep{2001MNRAS.322..891B}. We determined \teffpm{75}{5} and \loggpm{7.9}{0.3} for the WD of the system. Since we did not have optical spectra, the \logg error could only be assessed by Lyman lines. Our \teff and \logg values agree with those determined by \citet{2014MNRAS.440.1607B} and \citet{2018MNRAS.479.1612J}. On the other hand, only Si and S abundances derived by \citet{2014MNRAS.440.1607B} reside in our error range. We find C, P, and Fe abundances similar to theoretical predictions.
\smallskip
 
\item[\textbf{WD\,2353+026}] This DA WD has poor quality FUSE spectra. Considering the low metal content, distinguishing spectral lines from noise was not possible. Therefore, all abundances are given as upper limits. We derived \teffpm{61}{5} and \loggpm{7.5}{0.3} which agrees with the previous studies.

\end{description}


\onecolumn
\label{Appendix:Tables}
\section{Tables}

\begin{table*}[h!]
    \caption{Our WD sample and observations used in the spectral analysis.}
    \renewcommand*{\arraystretch}{1.25}
    \centering
    \begin{tabular}{l l l l l}
    \hline\hline
     WD Name$^{(a)}$ & Other Name & Program ID (FUSE) & Proposal ID (HST) & Optical$^{(b)}$\\
    \hline
     \multicolumn{5}{c}{DAO} \\
             & Longmore 1  & P133 & $-$  & UVES \\
    0439+466 & LS V +46 21 & M107 & 8638 & G10 \\     
    0500$-$156 & Abell 7   & B052 & $-$  & G10 \\     
    0615+556 & PuWe 1      & B052 & $-$  & G10 \\
    0823+316 & Ton 320     & B053 & $-$  & G10 \\     
    0834+500 & PG 0834+501 & B053 & $-$  & G10 \\
    0851+090 & Abell 31    & B052 & $-$  & UVES \\     
    1111+552 & NGC 3587    & B052 & $-$  & - \\     
    1214+267 & LB 2        & B053 & $-$  & G10 \\
    1253+378 & HZ 34       & B053 & $-$  & G10 \\     
    1957+225 & NGC 6853    & M107 & 8638 & G10 \\
    2226$-$210 & NGC 7293  & C177 & 8638 & G10 \\     
    2342+806 & GD 561      & B052 & 8329 & G10 \\ 
    \hline
    \multicolumn{5}{c}{DA} \\
    0027$-$636 & RE J0029$-$632 & Z903 & $-$  & G11 \\
    0229$-$481 & LB 1628        & M105 & $-$  & G11 \\
    0232+035 & Feige 24         & P104 & 7755 & G11 \\
    0311+480 & KPD 0311+4801    & Z904 & $-$  & G10 \\
    0343$-$007 & KUV03439$-$0048& E568 & $-$  & G11 \\
    0455$-$282 & RE J0457$-$280 & P104 & 14791& G10 \\
    0615+655 & HS 0615+6535     & E568 & 8636 & G10 \\
    0621$-$376 & RE J0623$-$374 & P104 & 14791& G10, UVES \\
    0939+262 & Ton 21           & $-$  & 8179 & G11 \\
    1056+516 & LB 1919          & D003 & $-$  & G11, SDSS \\
    1342+443 & PG 1342+444      & A034 & $-$  & G11, SDSS \\
    1738+669 & RE J1738+665     & M105 & 8179 & G10 \\
    1827+778 & HS 1827+7753     & E568 & $-$  & G11, SDSS \\ 
    2046+396 & KPD 2046+3940    & E568 & $-$  & G11 \\
    2146$-$433 & MCT2146$-$4320 & Z903 & $-$  & G10 \\        
    2211$-$495 & RE J2214$-$491 & M103 & 14791& G10, UVES \\    
    2218+706 & DeHt 5           & A034 & 8179 & G10 \\    
    2350$-$706 & HD 223816      & B120, A054  &$-$&$-$\\    
    2353+026 & PG 2353+027      & E568 & $-$  & G11 \\   

    \hline
    \end{tabular}  
    \smallskip
    \begin{tablenotes}
    \footnotesize
     {\item  \textbf{Notes.} $^{(a)}$ WD names from \citet{1999ApJS..121....1M}. $^{(b)}$ G10 and G11 refer to MWDD spectra used in \citet{2010ApJ...720..581G,2011ApJ...743..138G}, respectively. }
    \end{tablenotes}
    \label{tab:Observations}
\end{table*}

\begin{sidewaystable*}[]
\newcolumntype{e}[1]{D{+}{\,\pm\,}{#1}}
\small
\caption{\teff, \logg, and abundances (mass fractions) of DAO and DA WDs.}
\centering
\begin{tabular}{c e{2.1} r r r r r r r r r r r}
\hline\hline\noalign{\smallskip}
\multicolumn{1}{>{\centering\arraybackslash}m{1.5cm}}{(WD) Name} & \multicolumn{1}{>{\centering\arraybackslash}m{1cm}}{\teff [kK]} & \multicolumn{1}{>{\centering\arraybackslash}m{1.2cm}}{\logg  [cm/s$^2$]}  & \multicolumn{1}{>{\centering\arraybackslash}m{1.5cm}}{H}  & \multicolumn{1}{>{\centering\arraybackslash}m{1.5cm}}{He} & \multicolumn{1}{>{\centering\arraybackslash}m{1.5cm}}{C}  & \multicolumn{1}{>{\centering\arraybackslash}m{1.5cm}}{N}  & \multicolumn{1}{>{\centering\arraybackslash}m{1.5cm}}{O}  & \multicolumn{1}{>{\centering\arraybackslash}m{1.5cm}}{Si} & \multicolumn{1}{>{\centering\arraybackslash}m{1.5cm}}{P}  & \multicolumn{1}{>{\centering\arraybackslash}m{1.5cm}}{S}  & \multicolumn{1}{>{\centering\arraybackslash}m{1.5cm}}{Fe} & \multicolumn{1}{>{\centering\arraybackslash}m{1.5cm}}{Ni} \\
\hline\noalign{\smallskip}
\multicolumn{13}{c}{DAO} \\
Longmore 1  & 118 + 5 & 7.0 $\pm$ 0.3 & 6.93 $\times$ 10$\textsuperscript{-1}$ & 2.45 $\times$ 10$\textsuperscript{-1}$ & 1.74 $\times$ 10$\textsuperscript{-2}$ & 8.69 $\times$ 10$\textsuperscript{-3}$ & 3.29 $\times$ 10$\textsuperscript{-2}$ & 1.63 $\times$ 10$\textsuperscript{-4}$ & 5.81 $\times$ 10$\textsuperscript{-6}$ & 3.08 $\times$ 10$\textsuperscript{-4}$ & 2.40 $\times$ 10$\textsuperscript{-3}$ & \up6.78 $\times$ 10$\textsuperscript{-5}$ \\
0439+466 & 97 + 5  & 7.0 $\pm$ 0.2 & 9.57 $\times$ 10$\textsuperscript{-1}$ & 3.16 $\times$ 10$\textsuperscript{-2}$ & 1.15 $\times$ 10$\textsuperscript{-3}$ & 8.18 $\times$ 10$\textsuperscript{-4}$ & 2.69 $\times$ 10$\textsuperscript{-3}$ & 2.71 $\times$ 10$\textsuperscript{-4}$ & 3.79 $\times$ 10$\textsuperscript{-6}$ & 7.54 $\times$ 10$\textsuperscript{-5}$ & 6.20 $\times$ 10$\textsuperscript{-3}$ & 2.98 $\times$ 10$\textsuperscript{-4}$ \\
0500$-$156 & 104 + 6  & 7.2 $\pm$ 0.2 & 8.61 $\times$ 10$\textsuperscript{-1}$ & 1.25 $\times$ 10$\textsuperscript{-1}$ & 2.67 $\times$ 10$\textsuperscript{-3}$ & 2.95 $\times$ 10$\textsuperscript{-4}$ & 8.37 $\times$ 10$\textsuperscript{-3}$ & 1.65 $\times$ 10$\textsuperscript{-4}$ & 1.82 $\times$ 10$\textsuperscript{-6}$ & 3.90 $\times$ 10$\textsuperscript{-5}$ & 2.61 $\times$ 10$\textsuperscript{-3}$ & 1.48 $\times$ 10$\textsuperscript{-4}$ \\
0615+556 & 101 + 5 & 7.2 $\pm$ 0.2 & 9.35 $\times$ 10$\textsuperscript{-1}$ & 5.56 $\times$ 10$\textsuperscript{-2}$ & \up2.68 $\times$ 10$\textsuperscript{-3}$ & 2.95 $\times$ 10$\textsuperscript{-4}$ & 3.39 $\times$ 10$\textsuperscript{-3}$ & 1.42 $\times$ 10$\textsuperscript{-4}$ & 1.39 $\times$ 10$\textsuperscript{-5}$ & 1.19 $\times$ 10$\textsuperscript{-4}$ & 2.52 $\times$ 10$\textsuperscript{-3}$ & 2.49 $\times$ 10$\textsuperscript{-4}$ \\
0823+316 & 98 + 5  & 7.1 $\pm$ 0.2 & 9.75 $\times$ 10$\textsuperscript{-1}$ & 1.52 $\times$ 10$\textsuperscript{-2}$ & 2.77 $\times$ 10$\textsuperscript{-3}$ & 1.09 $\times$ 10$\textsuperscript{-3}$ & 2.55 $\times$ 10$\textsuperscript{-3}$ & 2.63 $\times$ 10$\textsuperscript{-4}$ & 5.81 $\times$ 10$\textsuperscript{-6}$ & 4.88 $\times$ 10$\textsuperscript{-5}$ & 2.70 $\times$ 10$\textsuperscript{-3}$ & 1.28 $\times$ 10$\textsuperscript{-4}$ \\
0834+500 & 90 + 3  & 7.0 $\pm$ 0.2 & 9.73 $\times$ 10$\textsuperscript{-1}$ & 1.82 $\times$ 10$\textsuperscript{-2}$ & 4.72 $\times$ 10$\textsuperscript{-4}$ & 1.84 $\times$ 10$\textsuperscript{-3}$ & 1.75 $\times$ 10$\textsuperscript{-3}$ & 1.22 $\times$ 10$\textsuperscript{-4}$ & 6.30 $\times$ 10$\textsuperscript{-6}$ & 7.44 $\times$ 10$\textsuperscript{-5}$ & 4.30 $\times$ 10$\textsuperscript{-3}$ & 3.47 $\times$ 10$\textsuperscript{-4}$ \\
0851+090 & 106 + 5 & 7.2 $\pm$ 0.2 & 8.60 $\times$ 10$\textsuperscript{-1}$ & 1.25 $\times$ 10$\textsuperscript{-1}$ & 2.87 $\times$ 10$\textsuperscript{-3}$ & 4.95 $\times$ 10$\textsuperscript{-4}$ & 8.37 $\times$ 10$\textsuperscript{-3}$ & 1.65 $\times$ 10$\textsuperscript{-4}$ & 5.82 $\times$ 10$\textsuperscript{-6}$ & 5.90 $\times$ 10$\textsuperscript{-5}$ & 2.51 $\times$ 10$\textsuperscript{-3}$ & 2.08 $\times$ 10$\textsuperscript{-4}$ \\
1111+552 & 110 + 5 & 7.1 $\pm$ 0.3 & 7.82 $\times$ 10$\textsuperscript{-1}$ & 2.07 $\times$ 10$\textsuperscript{-1}$ & 3.66 $\times$ 10$\textsuperscript{-3}$ & 4.25 $\times$ 10$\textsuperscript{-4}$ & 5.24 $\times$ 10$\textsuperscript{-3}$ & 1.22 $\times$ 10$\textsuperscript{-4}$ & 3.81 $\times$ 10$\textsuperscript{-6}$ & 4.87 $\times$ 10$\textsuperscript{-5}$ & 1.30 $\times$ 10$\textsuperscript{-3}$ & 1.30 $\times$ 10$\textsuperscript{-4}$ \\
1214+267 & 91 + 3  & 7.1 $\pm$ 0.3 & 9.69 $\times$ 10$\textsuperscript{-1}$ & 2.42 $\times$ 10$\textsuperscript{-2}$ & 6.73 $\times$ 10$\textsuperscript{-4}$ & 6.94 $\times$ 10$\textsuperscript{-4}$ & 1.75 $\times$ 10$\textsuperscript{-3}$ & 2.42 $\times$ 10$\textsuperscript{-4}$ & 6.20 $\times$ 10$\textsuperscript{-6}$ & 2.87 $\times$ 10$\textsuperscript{-5}$ & 1.93 $\times$ 10$\textsuperscript{-3}$ & 1.18 $\times$ 10$\textsuperscript{-4}$ \\
1253+378 & 93 + 4  & 7.0 $\pm$ 0.2 & 9.50 $\times$ 10$\textsuperscript{-1}$ & 4.07 $\times$ 10$\textsuperscript{-2}$ & 1.10 $\times$ 10$\textsuperscript{-3}$ & 7.22 $\times$ 10$\textsuperscript{-5}$ & 3.50 $\times$ 10$\textsuperscript{-3}$ & 5.00 $\times$ 10$\textsuperscript{-5}$ & 3.78 $\times$ 10$\textsuperscript{-6}$ & 3.72 $\times$ 10$\textsuperscript{-5}$ & 3.39 $\times$ 10$\textsuperscript{-3}$ & 1.18 $\times$ 10$\textsuperscript{-4}$ \\ 
1957+225 & 134 + 10 & 6.9 $\pm$ 0.4 & 7.39 $\times$ 10$\textsuperscript{-1}$ & 2.50 $\times$ 10$\textsuperscript{-1}$ & 2.36 $\times$ 10$\textsuperscript{-3}$ & 6.98 $\times$ 10$\textsuperscript{-4}$ & 5.77 $\times$ 10$\textsuperscript{-3}$ & 2.05 $\times$ 10$\textsuperscript{-4}$ & 5.55 $\times$ 10$\textsuperscript{-6}$ & 1.61 $\times$ 10$\textsuperscript{-4}$ & 1.21 $\times$ 10$\textsuperscript{-3}$ & \up6.80 $\times$ 10$\textsuperscript{-5}$ \\
2226$-$210 & 120 + 5 & 7.2 $\pm$ 0.3 & 8.80 $\times$ 10$\textsuperscript{-1}$ & 1.05 $\times$ 10$\textsuperscript{-1}$ & 2.36 $\times$ 10$\textsuperscript{-3}$ & 6.94 $\times$ 10$\textsuperscript{-4}$ & 1.93 $\times$ 10$\textsuperscript{-3}$ & \up2.03 $\times$ 10$\textsuperscript{-4}$ & 5.81 $\times$ 10$\textsuperscript{-6}$ & 1.08 $\times$ 10$\textsuperscript{-4}$ & 9.40 $\times$ 10$\textsuperscript{-3}$ & \up3.08 $\times$ 10$\textsuperscript{-4}$ \\
2342+806 & 83 + 5  & 7.2 $\pm$ 0.2 & 9.74 $\times$ 10$\textsuperscript{-1}$ & 1.63 $\times$ 10$\textsuperscript{-2}$ & 1.57 $\times$ 10$\textsuperscript{-3}$ & 1.94 $\times$ 10$\textsuperscript{-4}$ & 1.96 $\times$ 10$\textsuperscript{-3}$ & 2.43 $\times$ 10$\textsuperscript{-4}$ & 5.81 $\times$ 10$\textsuperscript{-6}$ & 1.09 $\times$ 10$\textsuperscript{-4}$ & 5.22 $\times$ 10$\textsuperscript{-3}$ & 3.88 $\times$ 10$\textsuperscript{-4}$ \\
\hline\noalign{\smallskip}
\multicolumn{13}{c}{DA} \\
0027$-$636 & 59 + 5 & 8.0 $\pm$ 0.2     &       9.99 $\times$ 10$\textsuperscript{-1}$  & \up3.98 $\times$ 10$\textsuperscript{-5}$ &     \up1.05 $\times$ 10$\textsuperscript{-6}$       &       \up5.04 $\times$ 10$\textsuperscript{-7}$       &       \up9.04 $\times$ 10$\textsuperscript{-6}$       &       3.32 $\times$ 10$\textsuperscript{-7}$       &       3.20 $\times$ 10$\textsuperscript{-8}$  &       \up4.30 $\times$ 10$\textsuperscript{-7}$       &       \up3.98 $\times$ 10$\textsuperscript{-4}$       &       \up3.49 $\times$ 10$\textsuperscript{-5}$       \\
0229$-$481 & 62 + 5 & 7.8 $\pm$ 0.2     &       9.99 $\times$ 10$\textsuperscript{-1}$  & \up1.98 $\times$ 10$\textsuperscript{-5}$ &     \up1.14 $\times$ 10$\textsuperscript{-6}$       &       \up1.08 $\times$ 10$\textsuperscript{-7}$       &       \up8.80 $\times$ 10$\textsuperscript{-7}$       &       9.22 $\times$ 10$\textsuperscript{-6}$       &       2.30 $\times$ 10$\textsuperscript{-6}$  &       4.98 $\times$ 10$\textsuperscript{-6}$       &       \up4.51 $\times$ 10$\textsuperscript{-4}$       &       \up3.49 $\times$ 10$\textsuperscript{-5}$       \\
0232+035 & 63 + 3 &     7.5 $\pm$ 0.2 & 9.99 $\times$ 10$\textsuperscript{-1}$  &       \up5.98 $\times$ 10$\textsuperscript{-5}$ &     1.45 $\times$ 10$\textsuperscript{-6}$  &       1.61 $\times$ 10$\textsuperscript{-6}$       &       1.04 $\times$ 10$\textsuperscript{-5}$  &       2.16 $\times$ 10$\textsuperscript{-5}$       &       3.14 $\times$ 10$\textsuperscript{-6}$  &       8.80 $\times$ 10$\textsuperscript{-6}$       &       6.20 $\times$ 10$\textsuperscript{-4}$  &       4.81 $\times$ 10$\textsuperscript{-5}$       \\
0311+480 & 70 + 10 & 7.3 $\pm$ 0.4      &       9.98 $\times$ 10$\textsuperscript{-1}$  & \up1.98 $\times$ 10$\textsuperscript{-5}$ &     5.55 $\times$ 10$\textsuperscript{-5}$  &       \up2.14 $\times$ 10$\textsuperscript{-4}$       &       \up1.11 $\times$ 10$\textsuperscript{-5}$       &       5.22 $\times$ 10$\textsuperscript{-6}$       &       \up1.50 $\times$ 10$\textsuperscript{-6}$       &       \up1.58 $\times$ 10$\textsuperscript{-5}$       &       1.80 $\times$ 10$\textsuperscript{-3}$  &       \up1.38 $\times$ 10$\textsuperscript{-4}$       \\
0343$-$007 & 63 + 4 & 7.7 $\pm$ 0.2     &       9.99 $\times$ 10$\textsuperscript{-1}$  & \up1.98 $\times$ 10$\textsuperscript{-5}$ &     \up1.85 $\times$ 10$\textsuperscript{-6}$       &       \up2.08 $\times$ 10$\textsuperscript{-6}$       &       \up8.10 $\times$ 10$\textsuperscript{-6}$       &       1.22 $\times$ 10$\textsuperscript{-5}$       &       \up3.50 $\times$ 10$\textsuperscript{-6}$       &       5.78 $\times$ 10$\textsuperscript{-6}$       &       \up6.91 $\times$ 10$\textsuperscript{-4}$       &       \up4.48 $\times$ 10$\textsuperscript{-5}$       \\
0455$-$282 & 66 + 3 & 7.5 $\pm$ 0.2     &       9.99 $\times$ 10$\textsuperscript{-1}$  & \up3.98 $\times$ 10$\textsuperscript{-5}$ &     \up1.45 $\times$ 10$\textsuperscript{-6}$       &       \up1.61 $\times$ 10$\textsuperscript{-6}$       &       5.69 $\times$ 10$\textsuperscript{-6}$  &       2.16 $\times$ 10$\textsuperscript{-5}$       &       3.14 $\times$ 10$\textsuperscript{-6}$  &       8.80 $\times$ 10$\textsuperscript{-6}$       &       5.20 $\times$ 10$\textsuperscript{-4}$  &       3.20 $\times$ 10$\textsuperscript{-5}$       \\
0615+655 & 83 + 10 & 7.6 $\pm$ 0.4      &       9.97 $\times$ 10$\textsuperscript{-1}$  &\up3.98 $\times$ 10$\textsuperscript{-5}$ &     1.45 $\times$ 10$\textsuperscript{-5}$  &       7.84 $\times$ 10$\textsuperscript{-6}$       &       3.61 $\times$ 10$\textsuperscript{-5}$  &       6.42 $\times$ 10$\textsuperscript{-5}$       &       \up5.20 $\times$ 10$\textsuperscript{-6}$       &       7.98 $\times$ 10$\textsuperscript{-5}$       &       2.51 $\times$ 10$\textsuperscript{-3}$  &       1.35 $\times$ 10$\textsuperscript{-4}$       \\
0621$-$376 & 65 + 3 & 7.5 $\pm$ 0.4 &   9.98 $\times$ 10$\textsuperscript{-1}$  &\up1.98 $\times$ 10$\textsuperscript{-5}$ &     1.06 $\times$ 10$\textsuperscript{-5}$  &       1.08 $\times$ 10$\textsuperscript{-5}$       &       3.22 $\times$ 10$\textsuperscript{-5}$  &       5.22 $\times$ 10$\textsuperscript{-5}$       &       4.20 $\times$ 10$\textsuperscript{-6}$  &       1.30 $\times$ 10$\textsuperscript{-5}$       &       1.72 $\times$ 10$\textsuperscript{-3}$  &       1.36 $\times$ 10$\textsuperscript{-4}$       \\
0939+262 & 66 + 3 &     7.7 $\pm$ 0.2  &        9.99 $\times$ 10$\textsuperscript{-1}$  & \up3.98 $\times$ 10$\textsuperscript{-5}$ &     1.45 $\times$ 10$\textsuperscript{-6}$  &       1.61 $\times$ 10$\textsuperscript{-6}$       &       3.69 $\times$ 10$\textsuperscript{-6}$  &       \up2.16 $\times$ 10$\textsuperscript{-5}$       &               &       8.80 $\times$ 10$\textsuperscript{-6}$        &       5.20 $\times$ 10$\textsuperscript{-4}$  &       3.40 $\times$ 10$\textsuperscript{-5}$       \\
1056+516 & 63 + 5 &     7.9 $\pm$ 0.2   &       9.99 $\times$ 10$\textsuperscript{-1}$  & \up1.98 $\times$ 10$\textsuperscript{-5}$ &     \up1.15 $\times$ 10$\textsuperscript{-6}$       &       \up5.84 $\times$ 10$\textsuperscript{-7}$       &       \up3.10 $\times$ 10$\textsuperscript{-6}$       &       \up1.11 $\times$ 10$\textsuperscript{-7}$       &       \up8.49 $\times$ 10$\textsuperscript{-9}$       &       \up7.77 $\times$ 10$\textsuperscript{-7}$       &       \up2.54 $\times$ 10$\textsuperscript{-4}$       &       \up2.89 $\times$ 10$\textsuperscript{-5}$       \\
1342+443 & 62 + 5 &     7.7 $\pm$ 0.3   &       9.99 $\times$ 10$\textsuperscript{-1}$  & \up1.98 $\times$ 10$\textsuperscript{-5}$ &     \up6.48 $\times$ 10$\textsuperscript{-6}$       &       \up2.08 $\times$ 10$\textsuperscript{-6}$       &       \up5.10 $\times$ 10$\textsuperscript{-5}$       &       6.22 $\times$ 10$\textsuperscript{-6}$       &       3.50 $\times$ 10$\textsuperscript{-6}$  &       1.28 $\times$ 10$\textsuperscript{-5}$       &       \up5.04 $\times$ 10$\textsuperscript{-4}$       &       \up3.88 $\times$ 10$\textsuperscript{-5}$       \\
1738+669 & 78 + 6 & 7.6 $\pm$ 0.4       &       9.99 $\times$ 10$\textsuperscript{-1}$  & \up9.18 $\times$ 10$\textsuperscript{-5}$ &     \up2.08 $\times$ 10$\textsuperscript{-6}$       &       \up2.04 $\times$ 10$\textsuperscript{-6}$       &       4.25 $\times$ 10$\textsuperscript{-6}$  &       1.84 $\times$ 10$\textsuperscript{-5}$       &       1.40 $\times$ 10$\textsuperscript{-6}$  &       1.78 $\times$ 10$\textsuperscript{-5}$       &       1.10 $\times$ 10$\textsuperscript{-3}$  &       6.09 $\times$ 10$\textsuperscript{-5}$       \\
1827+778 & 78 + 10 & 7.4 $\pm$ 0.4      &       9.98 $\times$ 10$\textsuperscript{-1}$  & \up1.98 $\times$ 10$\textsuperscript{-5}$ &     \up1.59 $\times$ 10$\textsuperscript{-4}$       &       \up1.74 $\times$ 10$\textsuperscript{-4}$       &       \up4.04 $\times$ 10$\textsuperscript{-6}$       &       3.62 $\times$ 10$\textsuperscript{-5}$       &       6.00 $\times$ 10$\textsuperscript{-6}$  &       \up1.16 $\times$ 10$\textsuperscript{-4}$       &       \up1.69 $\times$ 10$\textsuperscript{-3}$       &       \up1.03 $\times$ 10$\textsuperscript{-4}$       \\
2046+396 & 64 + 5 &     7.8 $\pm$ 0.3 & 9.99 $\times$ 10$\textsuperscript{-1}$  & \up1.98 $\times$ 10$\textsuperscript{-5}$ &     \up1.15 $\times$ 10$\textsuperscript{-6}$       &       \up4.84 $\times$ 10$\textsuperscript{-7}$       &       \up5.23 $\times$ 10$\textsuperscript{-7}$       &       7.21 $\times$ 10$\textsuperscript{-6}$       &       2.50 $\times$ 10$\textsuperscript{-6}$  &       5.17 $\times$ 10$\textsuperscript{-6}$       &       5.04 $\times$ 10$\textsuperscript{-4}$  &       \up4.64 $\times$ 10$\textsuperscript{-5}$       \\
2146$-$433 & 66 + 4 &7.5 $\pm$ 0.3 &    9.99 $\times$ 10$\textsuperscript{-1}$  & \up1.98 $\times$ 10$\textsuperscript{-5}$ &     6.15 $\times$ 10$\textsuperscript{-6}$  &       4.08 $\times$ 10$\textsuperscript{-6}$       &       \up4.23 $\times$ 10$\textsuperscript{-7}$       &       4.72 $\times$ 10$\textsuperscript{-5}$       &       7.50 $\times$ 10$\textsuperscript{-6}$  &       5.77 $\times$ 10$\textsuperscript{-6}$       &       1.23 $\times$ 10$\textsuperscript{-3}$  &       6.49 $\times$ 10$\textsuperscript{-5}$       \\
2211$-$495 & 68 + 4 & 7.4 $\pm$ 0.3     &       9.98 $\times$ 10$\textsuperscript{-1}$  & \up1.98 $\times$ 10$\textsuperscript{-5}$ &     1.35 $\times$ 10$\textsuperscript{-5}$  &       1.08 $\times$ 10$\textsuperscript{-5}$       &       2.63 $\times$ 10$\textsuperscript{-5}$  &       5.62 $\times$ 10$\textsuperscript{-5}$       &       7.49 $\times$ 10$\textsuperscript{-6}$  &       3.08 $\times$ 10$\textsuperscript{-5}$       &       1.61 $\times$ 10$\textsuperscript{-3}$  &       1.27 $\times$ 10$\textsuperscript{-4}$       \\
2218+706 & 78 + 5 &     7.4 $\pm$ 0.3   &       9.96 $\times$ 10$\textsuperscript{-1}$  & 5.10 $\times$ 10$\textsuperscript{-4}$ &        9.45 $\times$ 10$\textsuperscript{-5}$  &       1.12 $\times$ 10$\textsuperscript{-4}$       &       1.26 $\times$ 10$\textsuperscript{-4}$  &       6.42 $\times$ 10$\textsuperscript{-5}$       &       5.20 $\times$ 10$\textsuperscript{-6}$  &       4.58 $\times$ 10$\textsuperscript{-5}$       &       2.81 $\times$ 10$\textsuperscript{-3}$  &       2.15 $\times$ 10$\textsuperscript{-4}$       \\
2350$-$706 & 75 + 5 & 7.9 $\pm$ 0.3     &       9.99 $\times$ 10$\textsuperscript{-1}$  & \up1.98 $\times$ 10$\textsuperscript{-5}$ &     8.15 $\times$ 10$\textsuperscript{-6}$  &       1.08 $\times$ 10$\textsuperscript{-6}$       &       \up3.32 $\times$ 10$\textsuperscript{-7}$       &       5.62 $\times$ 10$\textsuperscript{-5}$       &       9.95 $\times$ 10$\textsuperscript{-6}$  &       1.88 $\times$ 10$\textsuperscript{-5}$       &       6.04 $\times$ 10$\textsuperscript{-4}$  &       \up4.58 $\times$ 10$\textsuperscript{-5}$       \\
2353+026 & 61 + 5 & 7.5 $\pm$ 0.3       &       9.99 $\times$ 10$\textsuperscript{-1}$  & \up1.98 $\times$ 10$\textsuperscript{-5}$ &     \up7.59 $\times$ 10$\textsuperscript{-6}$       &       \up1.08 $\times$ 10$\textsuperscript{-6}$       &       \up7.38 $\times$ 10$\textsuperscript{-7}$       &       \up1.97 $\times$ 10$\textsuperscript{-7}$       &       \up2.11 $\times$ 10$\textsuperscript{-8}$       &       \up3.08 $\times$ 10$\textsuperscript{-6}$       &       \up4.99 $\times$ 10$\textsuperscript{-4}$       &       \up2.48 $\times$ 10$\textsuperscript{-5}$       \\
\hline
\end{tabular}
\tablefoot{Upper limits are marked with triangles. Error limits for abundances do not exceed $\pm$ 0.5 dex.}
\label{tab:DAO_DA_abundances}
\end{sidewaystable*}

\begin{table*}[h!]
\caption{Previous analyses whose results are compared to the DAO WDs in our sample.}
\centering
    
\begin{tabular}{p{3cm} p{4cm} p{3cm} p{3cm} p{3cm}}
\hline\hline
Name & Optical & UV & Optical + UV & Abundances \\
\hline            
\multicolumn{5}{c}{DAO} \\

Longmore 1     & \comp{1}      & \comptwo{2}{3} &  $-$ & \comptwo{2}{3} \\
WD\,0439+466   & \compthr{4}{5}{6} & \comp{7}   & \comp{8}  & \comptwo{7}{8} \\ 
WD\,0500$-$156 & \compthrA{5}{6}{9} \compthr{10}{11}{12} & \comptwo{3}{12} & $-$ & \comptwo{3}{13} \\ 
WD\,0615+556   & \compthrA{5}{6}{11} \comp{12} & \comptwo{3}{12} & $-$ & \comptwo{3}{13} \\
WD\,0823+316   & \compthrA{4}{6}{12} \compthr{14}{15}{16} & \comp{12} & $-$ & \comp{13} \\     
WD\,0834+500   & \compthr{4}{6}{12} & \comp{12} & $-$ & \comp{13} \\
WD\,0851+090   & \comptwo{5}{12} & \comptwo{3}{12} & $-$ & \comptwo{3}{13} \\     
WD\,1111+552   & \comptwo{5}{11} & \comp{3} & $-$ & \comp{3} \\     
WD\,1214+267   & \compthrA{4}{6}{12} \comp{17} & \comp{12} & $-$ & \comp{13} \\
WD\,1253+378   & \compthrA{4}{5}{6} \compthr{10}{12}{18} & \comp{12} & $-$ & \comp{13} \\     
WD\,1957+225   & \comptwo{5}{6} & \comp{7} & \comp{3} & \comptwo{3}{7}\\
WD\,2226$-$210 & \compthrA{1}{5}{6} \comptwo{11}{19} & \comp{7} & \comp{3} & \comptwo{3}{7} \\     WD\,2342+806   & \compthrA{4}{5}{6} \comp{12} & \comptwo{3}{12} &  $-$ & \comptwo{3}{13}  \\ 
    \hline
\end{tabular} 
    \\
{\raggedright \textbf{References.} \comp{1}\citet{1985A&A...142..289M}, \comp{2}\citet{2004PASP..116..391H},  \comp{3}\citet{2012PhDT.......152Z}, \comp{4}\citet{1994ApJ...432..305B},  \comp{5}\citet{1999A&A...350..101N}, \comp{6}\citet{2010ApJ...720..581G},  \comp{7}\citet{2005ASPC..334..325T}, \comp{8}\citet{2007A&A...470..317R},  \comp{9}\citet{1981A&A...101..323M}, \comp{10}\citet{1985ApJS...58..379W}, \comp{11}\citet{1997IAUS..180..120M}, \comp{12}\citet{2004MNRAS.355.1031G}, \comp{13}\citet{2005MNRAS.363..183G}, \comp{14}\citet{2011ApJ...730..128T}, \comp{15}\citet{2019MNRAS.486.2169K}, \comp{16}\citet{2020ApJ...901...93B}, \comp{17}\citet{2015MNRAS.454.2787G}, \comp{18}\citet{2023ApJ...942..109L}, \comp{19}\citet{1988A&A...190..113M}. \par}
\tablefoot{The second and third columns list studies that only utilized optical and UV data, respectively, or the analyses derived separate parameters for both wavebands, whereas the fourth column displays the ones that reached a common \teff and \logg from both wavelength ranges. The fifth column lists works that performed abundance analyses.}
\label{tab:compareDAO}
\end{table*}

\begin{table*}[htp]
\caption{Like Table \ref{tab:compareDAO}, but for DAs and including EUV analyses as well.}
\begin{tabular}{p{2cm} p{4cm} p{3.5cm} p{1.5cm} p{2.1cm} p{2.5cm}}
\hline\hline
Name & Optical & UV &  EUV & Optical + UV & Abundances \\
\hline            
\\
WD\,0027$-$636 & \compthrA{1}{2}{3} \compthrA{4}{5}{6} \comptwo{7}{8} & \comptwo{5}{9} & \comp{10}  & $-$ & $-$  \\
WD\,0229$-$481 & \compthrA{1}{2}{6} \compthrA{7}{11}{12} \comp{13} & \comptwo{9}{13} & $-$ & $-$ &\comp{9}\\
WD\,0232+035 & \compthrA{1}{2}{3} \comptwoA{7}{13} \comptwo{14}{15} & \compthrA{5}{9}{13} \comptwoA{16}{17} \comptwo{18}{19}  & \comptwo{10}{20} & $-$ & \compthrA{9}{14}{17} \comp{18}\\
WD\,0311+480 & \comp{22} & $-$ & $-$ & $-$ & $-$  \\
WD\,0343$-$007 & \compthrA{1}{6}{7} \compthr{11}{13}{23} & \comp{13} & $-$ & $-$ & $-$ \\
WD\,0455$-$282 & \compthrA{1}{2}{3} \compthrA{5}{6}{8} \compthrA{12}{13}{14} \compthr{22}{24}{25} & \compthrA{5}{9}{21} \compthr{24}{26}{27} & \comp{10} & \comp{24} & \compthrA{9}{14}{26} \comp{27}\\
WD\,0615+655   & \compthrA{13}{22}{28} & \comp{13} & $-$ & $-$ & $-$ \\
WD\,0621$-$376 & \compthrA{1}{2}{3} \compthrA{5}{6}{13} \compthr{14}{22}{25} & \compthrA{5}{9}{13} \comptwo{21}{27} & \comp{10} & \comp{29} & \compthrA{9}{14}{27} \comp{29} \\
WD\,0939+262 &  \compthrA{1}{7}{13} \compthr{14}{30}{31}  & \comp{13} & $-$ & $-$ & \comp{14} \\
WD\,1056+516 &  \compthrA{3}{7}{13} \compthrA{32}{33}{34} \comp{35} & \comptwo{13}{36} & \comp{10} \comp{36} &  $-$ & \comp{36}  \\
WD\,1342+443 & \compthrA{7}{14}{30} \compthrA{31}{33}{34} \comp{35} & \comptwo{9}{14} & $-$ & \comp{37} & \compthr{9}{14}{37} \\
WD\,1738+669 & \compthrA{1}{2}{13} \comptwo{14}{22} & \comptwo{13}{21} & $-$ & $-$ & \comp{14} \\
WD\,1827+778 & \compthrA{7}{28}{33} \comptwo{34}{35} & $-$ & $-$ & $-$ & $-$ \\ 
WD\,2046+396 & \compthrA{1}{7}{13} \comp{38} & \comp{13} & $-$ & $-$ & $-$ \\
WD\,2146$-$433 & \compthrA{1}{4}{6} \compthr{8}{13}{22} & \comptwo{9}{13} & $-$ & $-$ & \comp{9} \\     
WD\,2211$-$495 & \compthrA{1}{2}{3} \compthrA{5}{6}{12} \compthr{14}{22}{25} & \compthrA{5}{9}{21} \comp{27} & \comp{10} & \comp{29} & \compthrA{9}{14}{27} \comp{29} \\    
WD\,2218+706 & \compthrA{14}{22}{39} \comp{40} & \comp{40} & $-$ & \comp{41} & \comptwo{14}{42} \\    
WD\,2350$-$706 & $-$ & \comptwo{9}{14} & $-$ & $-$ & \comp{9} \\    
WD\,2353+026 & \compthrA{1}{4}{6} \compthrA{7}{13}{28} \comp{31} & \comp{13} & $-$ & $-$ & $-$  \\   
\hline
\end{tabular} 
    \\
{\raggedright \textbf{References.}\comp{1}\citet{1997ApJ...488..375F}, \comp{2}\citet{1997MNRAS.286..369M}, \comp{3}\citet{1997ApJ...480..714V},  \comp{4}\citet{2005A&A...432.1025K}, \comp{5}\citet{2005ASPC..334..185V}, \comp{6}\citet{2009A&A...505..441K}, \comp{7}\citet{2011ApJ...743..138G}, \comp{8}\citet{2021AJ....162..188B}, \comp{9}\citet{2014MNRAS.440.1607B}, \comp{10}\citet{2002A&A...382..164S}, \comp{11}\citet{1995ApJ...443..735B}, \comp{12}\citet{2001A&A...378..556K}, \comp{13}\citet{2007ApJ...667.1126L}, \comp{14}\citet{2003MNRAS.341..870B}, \comp{15}\citet{2017ApJ...848...11B}, \comp{16}\citet{1986ApJ...306..629H}, \comp{17}\citet{1992ApJ...392L..27V}, \comp{18}\citet{2000ApJ...544..423V}, \comp{19}\citet{2018MNRAS.479.1612J}, \comp{20}\citet{1989ApJ...336L..25V}, \comp{21}\citet{2003MNRAS.344..562B}, \comp{22}\citet{2010ApJ...720..581G}, \comp{23}\citet{2010ApJ...714.1037L}, \comp{24}\citet{1998MNRAS.299..520B}, \comp{25}\citet{2007ApJ...654..499K}, \comp{26}\citet{1996ApJ...468..898V}, \comp{27}\citet{2019MNRAS.487.3470P}, \comp{28}\citet{1998A&A...338..563H}, \comp{29}\citet{1993ApJ...416..806H}, \comp{30}\citet{1994ApJ...432..305B}, \comp{31}\citet{2005ApJS..156...47L}, \comp{32}\citet{2011ApJ...730..128T}, \comp{33}\citet{2019MNRAS.486.2169K}, \comp{34}\citet{2019MNRAS.482.5222T}, \comp{35}\citet{2020ApJ...901...93B}, \comp{36}\citet{2012A&A...546A...1A}, \comp{37}\citet{2002MNRAS.330..425B}, \comp{38}\citet{2011AJ....141...96L}, \comp{39}\citet{1999A&A...350..101N}, \comp{40}\citet{2004MNRAS.355.1031G}, \comp{41}\citet{2001MNRAS.325.1149B}, \comp{42}\citet{2005MNRAS.363..183G} 

\par}

\label{tab:compareDA}
\end{table*}
\FloatBarrier

\end{appendix}

\end{document}